\newlength{\absize}
\documentclass[12pt]{article}

\usepackage{psfig}
\usepackage{epsfig}
\usepackage{amssymb}

\renewcommand{\baselinestretch}{1.5}

\begin{document}
\thispagestyle{empty}
\pagestyle{empty}
\renewcommand{\thefootnote}{\fnsymbol{footnote}}
\newcommand{\starttext}{\newpage\normalsize
\pagestyle{plain}
\setlength{\baselineskip}{3ex}\par
\setcounter{footnote}{0}
\renewcommand{\thefootnote}{\arabic{footnote}}
}

\newcommand{\preprint}[1]{\begin{flushright}
\setlength{\baselineskip}{3ex}#1\end{flushright}}
\renewcommand{\title}[1]{\begin{center}\LARGE
#1\end{center}\par}
\renewcommand{\author}[1]{\vspace{2ex}{\Large\begin{center}
\setlength{\baselineskip}{3ex}#1\par\end{center}}}
\renewcommand{\thanks}[1]{\footnote{#1}}
\renewcommand{\abstract}[1]{\vspace{2ex}\normalsize\begin{center}
\centerline{\bf Abstract}\par\vspace{2ex}\parbox{\absize}{#1
\setlength{\baselineskip}{2.5ex}\par}
\end{center}}

\newlength{\eqnparsize}
\setlength{\eqnparsize}{.95\textwidth}

\setlength{\jot}{1.5ex}
\newcommand{\figsize}{\small}
\renewcommand{\bar}{\overline}
\font\fiverm=cmr5
\input prepictex
\input pictex
\input postpictex
\input{psfig.sty}
\newdimen\tdim
\tdim=\unitlength

\setcounter{bottomnumber}{2}
\setcounter{topnumber}{3}
\setcounter{totalnumber}{4}
\renewcommand{\bottomfraction}{1}
\renewcommand{\topfraction}{1}
\renewcommand{\textfraction}{0}

\def\draft{
\renewcommand{\label}[1]{{\quad[\sf ##1]}}
\renewcommand{\ref}[1]{{[\sf ##1]}}
\renewenvironment{thebibliography}{\section*{References}}{}
\renewcommand{\cite}[1]{{\sf[##1]}}
\renewcommand{\bibitem}[1]{\par\noindent{\sf[##1]}}
}

\def\theequation{\thesection.\arabic{equation}}
\preprint{\#HUTP-02/A043\\ 9/02}
\title{ $\mathcal{N}=1$ Supersymmetric $SU(2)^{r}$ Moose Theories\thanks{
Research supported in part by the National Science Foundation under
grant number NSF-PHY/98-02709.
}}
\author{
Girma Hailu\thanks{hailu@feynman.harvard.edu}
\\
\small\sl Lyman Laboratory of Physics \\
\small\sl Harvard University \\
\small\sl Cambridge, MA 02138
}


\abstract{
We study the quantum moduli spaces and dynamical superpotentials of four dimensional $SU(2)^r$
linear and ring  moose theories with $\mathcal{N}=1$ supersymmetry and
link chiral superfields in the fundamental representation.
Nontrivial quantum moduli spaces and dynamical superpotentials
are produced.
When the moduli space is perturbed by a generic tree level superpotential,
the vacuum space becomes discrete. 
The ring moose is in the Coulomb phase and we find two singular submanifolds with a 
nontrivial modulus that is a function of   
 all the independent gauge invariants needed to parameterize the quantum moduli space. 
The massive theory near these singularities confines.
The Seiberg-Witten elliptic curve that describes the quantum moduli space
of the ring moose is produced. 
 }

\starttext


\setcounter{equation}{0}
\section{{\label{sec:Intr}Introduction}}

There are good motivations to study four dimensional
moose \cite{G-1} (or quiver \cite{douglas-moore-1}) theories. On one
hand, a class of these theories has been shown to give a description
of extra dimensions \cite{ACG-1,HPW-1}. Consequently, extra dimensions
can be naturally incorporated within a familiar setting of four dimensional
gauge theories. What is amusing in this {}``deconstruction'' of
extra dimensions is that the extra dimensions could be generated with
few number of nodes and links and the {}``size'' between the nodes
gives a UV completion of the four dimensional gauge theory. Furthermore,
deconstruction has provided a framework for model building and investigating
various issues such as electroweak symmetry breaking and accelerated
grand unification \cite{ACG-2}. On the other hand, from a different
direction, the supersymmetric versions of similar moose diagrams appear
in type IIA string theory with D6 branes wrapped on $S^{3}$ of Calabi-Yau
threefold of $T^{*}S^{3}$ and also in type IIB string theory
with D3, D5 and D7 branes wrapped over various cycles of Calabi-Yau
threefold. See \cite{CFIKV} for a recent discussion on this.

Moose diagrams contain nodes and links. Each node represents a gauge
group and each link represents a matter field that transforms as some
nontrivial representation of the gauge groups directly linked to it and as
singlet under the rest. The original motivation
for moose diagrams was to give a succinct graphical representation
for encoding the transformations of fermions under various gauge (and
global) symmetries. The transformation of a moose diagram into a description
of extra dimensions occurs when the link fields develop vacuum expectation
value (VEV) and ``hop'' between the nodes. It has been well know for
sometime that four dimensional supersymmetric gauge theories have
classical moduli space of vacua. If quantum fluctuations
do not give rise to a non-vanishing dynamical superpotential, the
quantum theory will have a quantum moduli space of vacua \cite{seiberg-1}.
In fact, supersymmetric gauge theories have larger moduli
spaces of vacua than non-supersymmetric theories. Therefore, supersymmetric
moose theories could furnish a richer framework for model building
based on the idea of deconstruction.

In this note we are interested in $\mathcal{N}=1$ supersymmetric
$SU(2)^{r}$ linear and ring moose theories where the gauge group
at each node is $SU(2)$ and the links are chiral superfields that
transform as fundamentals under the nearest gauge groups and as
singlets under the rest. The linear moose will look like $A_{\mathrm{r }}$
Dynkin diagram with additional link fields at the ends. We will call
a chiral superfield $Q_{i}$ that links two nodes $SU(2)_{i}$ and
$SU(2)_{i+1}$ internal and $Q_{i}$ transforms as $(\square,\,\square)$ under
$SU(2)_{i}\times SU(2)_{i+1}$. We will call the superfields $Q_{0}$
and $Q_{r}$ at the ends of a linear moose external. The external link 
$Q_{0}$ transforms as $\square$ under $SU(2)_{1}$ and $Q_{r}$ transforms
as $\square$ under $SU(2)_{r}$. The internal links are doublets carrying
two $SU(2)$ colors indices while the external links are each two
doublets with $SU(2)$ subflavor symmetry and they carry one color
and one subflavor indices. The ring moose will look like affine $\hat{A}_{r}$
Dynkin diagram with all links carrying two color
indices. Both the linear and the ring moose theories are asymptotically
free and anomaly free.

We will obtain nontrivial quantum moduli
space constraints and dynamical  superpotentials starting from simple pure gauge theories of
disconnected nodes by exploiting simple and
efficient integrating in \cite{ILS-1,Intriligator-1}  and out procedures. 
We will find that the linear moose with
both external links present has a
quantum moduli space of vacua.  Explicit
parameterization of the vacua in terms of the gauge invariant
objects constructed out of the chiral superfields will be found. We will also study the
vacuum structure of this theory for a specific case of a moose with two nodes when
perturbed by a tree level superpotential that includes a non-quadratic 
gauge singlet. We will find that this leads to a discrete vacuum space.
The linear mooses
without one or both external links have non-vanishing
dynamical superpotentials and we will explicitly compute these
superpotentials. 
A generic point in the moduli space of the ring moose has an unbroken $U(1)$
gauge symmetry and the ring moose is in the Coulomb phase.    
We will find two singular submanifolds with modulus that is a nontrivial function 
of all the independent gauge invariant objects
needed to parameterize the moduli space of the ring moose.
The massive theory near these singularities leads to confinement. 
The Seiberg-Witten elliptic curve that describes the
quantum moduli space of the ring moose will follow from our computation. 

The Seiberg-Witten elliptic curve of the ring moose was computed in \cite{CEFS} using a different
method where it was started with the curve for a ring with two nodes given in  
\cite{IS-1}  and  various asymptotic limits and symmetry arguments were used 
to obtain the curve for a ring moose with three nodes. The result was then
generalized to the curve for a ring with arbitrary number of nodes. Here we will directly and 
explicitly compute the singularities of the quantum moduli space and the corresponding 
Seiberg-Witten elliptic curve for a ring moose with arbitrary number of nodes. 
The curve we obtain agrees with \cite{IS-1} for a ring moose with two nodes and with \cite{CEFS} for a ring 
moose with three nodes. We believe that the curve given in \cite{CEFS} is incorrect for ring mooses with 
four or more nodes. 


\setcounter{equation}{0}
\section{{\label{sec:int-in-out}Integrating in}}

In this section we will briefly summarize the integrating in 
procedure of \cite{ILS-1,Intriligator-1} in the context of the moose theories we will be
studying. 
Consider
$\mathcal{N}=1$ supersymmetric gauge theory with gauge group $G=\prod _{i=1}^{r}SU(2)_{i}$
and matter chiral superfield $Q_{i}$ transforming as $\square$  under
the gauge groups that are directly  linked to it.
The parameters we need to describe the dynamical superpotential of
this theory are gauge singlet fields $X_{j}$ constructed out
of $Q_{i}$ and the nonperturbative dynamical scale $\Lambda _{i}$
of each $SU(2)_{i}$. Let us denote this superpotential by $W_{\mathrm{u}}(X_{j},\, \Lambda _{i})$.
Now suppose we give mass $m_{k}$ to one of the chiral superfields $Q_{k}$. For energies below
$m_{k}$, we integrate out $Q_{k}$.  This can be achieved by  integrating out
all gauge singlets that contain
$Q_{k}$. All gauge invariant objects that contain $Q_{k}$ will then be absent in the lower
energy theory.
If we denote those gauge singlets in $X_{j}$ that do
not contain $Q_{k}$ by $Y_{j}$ and those that contain $Q_{k}$ by
$Z_{j}$, then the dynamical superpotential of the lower energy theory can be
written as $W_{\mathrm{d}}(Y_{j},\, \Lambda _{id})$, where
$\Lambda _{id}$ is the nonperturbative dynamical scale of
$SU(2)_{i}$ in the lower energy theory. The integrating in 
procedure takes us from  $W_{\mathrm{d}}$ to
$W_{\mathrm{u}}$.

First suppose we know $W_{\mathrm{u}}(X_{j},\, \Lambda _{i})$. In
order to compute $W_{\mathrm{d}}(Y_{j},\, \Lambda _{id})$, first
we add the tree level superpotential
\begin{equation}
W_{\mathrm{tree}}={\sum_{j}}{g_{j}Z_{j}} \label{eq:wtree1}
\end{equation}
to $W_{\mathrm{u}}(X_{j},\, \Lambda _{i})$, where $g_{j}$ are coupling
constants. One of the terms in $W_{\mathrm{tree }}$ is $m_{k}M_{k}$,
where $M_{k}=\mathrm{det}\,(Q_{k})$ is a quadratic gauge singlet. This term
gives mass $m_{k}$ to $Q_{k}$. In general, $Z_{j}$ also consists of gauge
singlets that are not quadratic
in $Q_{k}$. Integrating out $Q_{k}$ in
\begin{equation}
W=W_{\mathrm{u}}+W_{\mathrm{tree}}\label{eq:inter-1a}
\end{equation}
gives
\begin{equation}
W=W_{\mathrm{d}}+W_{\mathrm{tree,d}}+W_{\Delta},\label{eq:inter-1b}
\end{equation}
where
\begin{equation}
W_{\mathrm{tree,d}}=W_{\mathrm{tree}}|_{\langle Q_{k}\rangle}.\label{eq:inter-1c}
\end{equation}
We will see in Section  (\ref{sec-linear-su2r})
that $W_{\Delta}=0$ in all the mooses we will be studying.

Suppose we know $W_{\mathrm{d}}(X_{j},\, \Lambda _{id})$ instead.
Integrating in $Q_{k}$ is equivalent to making a Legendre
transformation from $W_{\mathrm{d}}(Y_{j},\Lambda _{id})$ to $W_{\mathrm{u}}(X_{j},\, \Lambda _{i})$.
Matching the high energy and low energy scales $\Lambda _{i}$ and
$\Lambda _{id}$ at $m_{k}$, $\Lambda _{id}$ can
be expressed in terms of $\Lambda _{i}$ and $m_{k}$. Let us write
$\Lambda _{id}=\Lambda _{i}(m_{k})$. The higher energy dynamical superpotential 
$W_{\mathrm{u}}(X_{j},\, \Lambda _{i})$ is then obtained by integrating out $g_{j}$
(which consists of $m_{k}$) in
\begin{equation}
W=W_{\mathrm{d}}(\mathrm{with }\,\Lambda _{id}\rightarrow \Lambda _{i}(m_{k}))\mathrm{ }+
W_{\mathrm{tree,d}}-W_{\mathrm{tree}}.\label{eq:inter-2a}
\end{equation}


\setcounter{equation}{0}
\section{\label{sec-linear-su2r}Linear moose\protect
\footnote{Quantum moduli
space constraint relations
for a linear moose with two and more nodes were first shown to us by Howard Georgi. Many results in
this section overlap with results in \cite{S-H}.
}}

In this section we study the quantum moduli space of a linear
moose of $\mathcal{N}=1$ supersymmetric  $SU(2)^{r}$ gauge theory.
An internal chiral superfield
$Q_{i}$ links the $i^{th}$ and $(i+1)^{th}$ nodes. The internal
link $Q_{i}$
transforms as $(\square,\,\square)$ under $SU(2)_{i}\times SU(2)_{i+1}$ and as
singlet under all the other gauge groups. One of the external
links $Q_{0}$ transforms as $\square$ under $SU(2)_{1}$ and the second external link 
$Q_{r}$ transforms as $\square$ under $SU(2)_{r}$.
Each external link is two doublets with $SU(2)$
subflavor symmetry. We will compute the quantum moduli space of this theory
starting from pure disconnected
gauge groups and integrating in all the link fields. Gaugino
condensation in the pure gauge theory gives
a nonperturbative superpotential,\begin{equation}
W=\sum _{i=1}^{r}2\epsilon _{i}\Lambda _{0i}^{3},\label{sp-1}
\end{equation}
where each $\epsilon _{i}=\pm 1$ labels the two
vacua due to the breaking of the $Z_{4}$ $R$ symmetry to $Z_{2}$
 and $\Lambda _{0i}$ is the
nonperturbative dynamical scale of $SU(2)_{i}$.  We will no more use ``d'' and ``u''
subscripts in $W$ as it should be obvious in all cases.
Our notation for the dynamical scales is $\Lambda
_{0i}$ for the scale of $SU(2)_{i}$ with no matter linked,
$\Lambda _{id}$ when there is one link, and $\Lambda
_{i}$ when there are two links attached.
The scale $\Lambda _{i}$  is related to $\Lambda _{0i}$ by threshold matching of
the gauge coupling running at the masses $m_{i-1}$ and $m_{i}$ of
$Q_{i-1}$ and $Q_{i}$ respectively,
\begin{equation}
\Lambda _{0i}^{6}=\Lambda _{i}^{4}m_{i-1}m_{i}.\label{scales-1}
\end{equation}

In order to appreciate the power and simplicity of the integrating
in procedure in producing quantum moduli space constraints and exact dynamical superpotentials, 
we will start
with building up the chains
shown in Figures \ref{fig-lmoose1}(b) - \ref{fig-lmoose1}(e).
{\figsize\begin{figure}[htb]
{\centerline{\epsfxsize=8cm \epsfbox{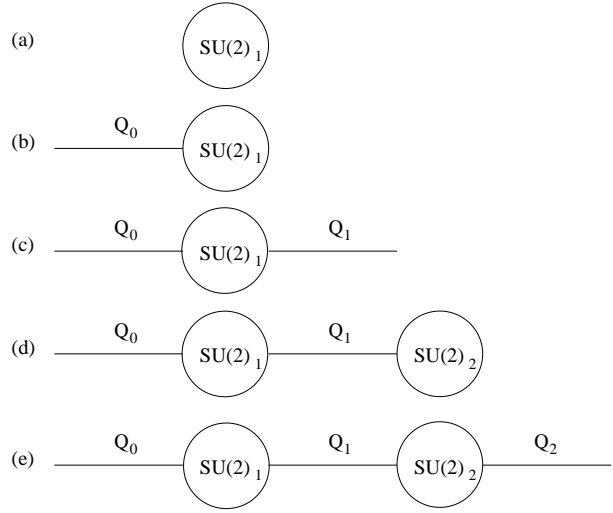}}}
\caption{\figsize\sf \label{fig-lmoose1}
Linear moose with (a) One node and no link, (b)
One node and one link, (c) One node and two links, (d) Two nodes
and one internal and one external links, and (e) Two nodes and three
links.  The external links  each have
one color and one subflavor indices and each internal link has
two color indices.
}\end{figure}}
Later, we will directly
and more formally compute the moduli space constraint for a general case of a linear
moose with arbitrary number of nodes.

First let us integrate in
$(Q_{0})_{f\beta_{0} }$, where we have explicitly put the subflavor
$f=1,\, 2$ and color $\beta_{0} =1,\, 2$ indices of $Q_{0}$, and 
build Figure \ref{fig-lmoose1}(b) starting from  Figure \ref{fig-lmoose1}(a).
We use indices $\alpha_{i}$, $\beta_{i}$ for color and indices $f$, $g$ for subflavor.
The integrating in procedure in this case is simple. There is
only one gauge
and flavor invariant mass term given by  $W_{\mathrm{tree}}=m_{0}M_{0}$,
where $M_{0}=\frac{1}{2}(Q_{0})_{f \beta _{0}}(Q_{0})_{f' {\beta}' _{0}}\epsilon ^{ff'} \epsilon
^{\beta _{0}{\beta}' _{0}} = \mathrm{det}\,(Q_{0})$ and $m_{0}$ is
a constant. Threshold matching at energy $m_{0}$ gives $\Lambda_{01}^{3}=(m_{0}\Lambda_{1d}^{5})^{1/2}$.
To integrate in $Q_{0}$, first we 
replace $\Lambda_{01}^{3}\rightarrow (m_{0}\Lambda_{1d}^{5})^{1/2}$ and subtract $m_{0}M_{0}$ in
(\ref{sp-1}) for $r=1$,
\begin{equation}
W=2\epsilon _{1}\,(m_{0}\Lambda _{1d}^{5})^\frac{1}{2}-m_{0}M_{0}. \label{eq:su21upminim}
\end{equation}
We then minimize (\ref{eq:su21upminim}) with $m_{0}$ to obtain
\begin{equation}
W=\frac{\Lambda _{1d}^{5}}{M_{0}}.\label{eq:su21up}
\end{equation}
This is exactly the Afflek-Dine-Seiberg \cite{ADS} superpotential
of $SU(2)$ with one flavor coming from a single instanton in the completely broken $SU(2)_{1}$.
We can go back to the pure gauge theory by integrating out $M_{0}$ in
\begin{equation}
W=\frac{\Lambda _{1d}^{5}}{M_{0}}+m_{0}M_{0} \label{eq:su21back1}
\end{equation}
which gives the original superpotential (\ref{sp-1}) for $r=1$.

Next let us add a second external link $Q_{1}$ and build the moose diagram with one node and two links 
shown in Figure \ref{fig-lmoose1}(c). 
In addition to $M_{0}$, there are five more gauge
invariants given by $M_{1}=\frac{1}{2}(Q_{1})_{\alpha _{1} g}
(Q_{1})_{\alpha' _{1} g'}\epsilon ^{\alpha _{1}\alpha' _{1}}\epsilon ^{g g'}=\mathrm{det}
(Q_{1})$
and a $2 \times 2$ matrix $T$ with components $(T)_{f g}=(Q_{0})_{f
\beta _{0}}(Q_{1})_{\alpha _{1} g}\epsilon
^{\beta _{0}\alpha _{1}}$. Now $T$ is a non-quadratic gauge singlet and
we need to compute $W_{\mathrm{tree,d}}$. 
We can get to Figure \ref{fig-lmoose1}(c) either from Figure \ref{fig-lmoose1}(b) or
directly from Figure \ref{fig-lmoose1}(a). First let us go from  Figure \ref{fig-lmoose1}(b). 
The new gauge singlets  are $M_{1}$ and $T$.  In this case, $W_{\mathrm{tree,d}}$ is easily
computed by integrating out $Q_{1}$ in  $W_{\mathrm{tree}}=\mathrm{tr}\,(c\,T)+m_{1}M_{1}$,
where $c$ is a constant $2 \times 2$ matrix,
which gives $W_{\mathrm{tree,d}}=-\mathrm{det}\,(c)\, M_{0}/m_{1}$. 
Let us now show that symmetries and asymptotic limits give $W_{\Delta }=0$ in (\ref{eq:inter-1b})
for this example. $W_{\Delta }$ can only be a function of $M_{0}$, $\Lambda _{1d}^{5}$, $m_{1}$ and  $c$.
Moreover, there is
a $U(1)_{Q_{0}}\times U(1)_{Q_{1}}\times U(1)_{R_{1}}$
symmetry and  $M_{0}$, $\Lambda _{1d}^{5}$,  $m_{1}$ and $c$ have the following $(Q_{0},\, Q_{1},\, R _{1})$
charges: $M_{0}:$ $(2,\, 0,\, 0)$, $\Lambda _{1d}^{5}:$ $(2,\, 0,\, 2)$,
$m_{1}:$ $(0,\, -2,\, 2)$ and $c:$ $(-1,\, -1,\, 2)$. Since $W_{\Delta }$
must have charges $(0,\, 0,\, 2)$, the most general $W_{\Delta }$
is given by
\begin{equation}
W_{\Delta } = - \frac{\mathrm{det}\,(c)M_{0}}{m_{1}}f(t), \quad \mathrm{where}\quad t\equiv\frac{m_{1}
\Lambda _{1d}^{5}}{M_{0}^{2}\,\mathrm{det}\,(c)} \label{eq:wdelta}
\end{equation}
and the argument $t$ has charge $(0,\, 0,\, 0)$.
Since any dependence of  $W_{\Delta }$ on $\Lambda_{1d}^{5}$ can only come from instantons in
the completely broken $SU(2)_{1}$,
we can expand $f(t)$ as $f(t)= \sum_{n=1}^{\infty} a_{n} t^{n}$.
The case $n=0$ would have simply reproduced $W_{\mathrm{tree,d}}$.
On the other hand, $W_{\Delta }$ should obey the limits
\begin{equation}
\lim _{\Lambda _{1d} \rightarrow 0}W_{\Delta }=0,\quad
\lim _{m_{1} \rightarrow \infty }W_{\Delta }=0. \label{deltalimit-1}
\end{equation}
That is because when $\Lambda _{1d} \rightarrow 0$, the quantum superpotential reduces to the classical
superpotential with only $W_{\mathrm{tree,d}}$. Furthermore, in the limit  $m_{1} \rightarrow \infty$,
$Q_{1}$ should completely decouple from the low energy superpotential except for its effect on the scale of
the lower energy theory.
It follows from (\ref{eq:wdelta}), (\ref{deltalimit-1}) and the above expansion of $f(t)$ that $W_{\Delta }=0$. 
Similar arguments can be used to show that $W_{\Delta }=0$ for all
the moose diagrams we consider in this note, and we will not talk about
$W_{\Delta }$ any further.
We then minimize
\begin{equation}
W=\frac{\Lambda _{1d}^{5}}{M_{0}}
-\frac{\mathrm{det}\,(c)\, M_{0}}{m_{1}}-m_{1}M_{1}
-\mathrm{tr}\,(c\,T).\label{eq:su22-c01a}
\end{equation}
with $m_{1}$ and $c$  to obtain $W=0$ and
a moduli space of vacua with constraint \begin{equation}
\mathrm{det}\,T-M_{0}M_{1}+\Lambda _{1}^{4}=0.\label{eq:su21m-a}\end{equation} 

Now let us go directly from Figure \ref{fig-lmoose1}(a) to \ref{fig-lmoose1}(c).   
Which and how many chiral superfields do we need to integrate out of $\mathrm{tr}\,(c\,T)$  to 
compute $W_{\mathrm{tree,d}}$ in this case?
We will need to integrate out only one and either one of $Q_{0}$ or $Q_{1}$ will do the job. 
We can
look at this by thinking in terms of building a linear moose chain that has both external links. Such a linear moose has 
one non-quadratic $2\times2$ matrix gauge singlet. This non-quadratic gauge singlet disappears if any one of the link 
fields is removed. 
If we think in terms of building the whole moose chain by putting in a link at 
a time,
the need for $W_{\mathrm{tree,d}}$ arises only when we put in the last link where the 
non-quadratic gauge singlet comes in. 
For the current example, let us first choose to integrate out $Q_{1}$. This is done by 
minimizing the tree level superpotential $\mathrm{tr}\,(c\,T)+m_{1}M_{1}$ with $Q_{1}$ which 
gives $W_{\mathrm{tree,d}}=-\mathrm{det}\,(c)\, M_{0}/m_{1}$. 
In fact, once we have added the $W_{\mathrm{tree,d}}$ we obtain in this way  
to the superpotential as in (\ref{eq:inter-2a}), we can integrate in all the independent gauge invariants at 
the same time.
The superpotential we need for integrating in the two flavors is then
\begin{equation}
W=2\epsilon _{1}\,(m_{0}m_{1}\Lambda _{1}^{4})^{1/2}-\frac{\mathrm{det}\,(c)\, M_{0}}{m_{1}}-m_{0}M_{0}-m_{1}M_{1}
-\mathrm{tr}\,(c\,T).\label{eq:su22-c01}
\end{equation}
Minimizing this with $m_{0}$, $m_{1}$ and $c$ gives $W=0$  and (\ref{eq:su21m-a}).
If we had chosen to compute  $W_{\mathrm{tree,d}}$  by integrating out $Q_{0}$ in  $\mathrm{tr}\,(c\,T)+m_{0}M_{0}$ instead, 
we would have obtained   $W_{\mathrm{tree,d}}=-\mathrm{det}\,(c)\, M_{1}/m_{0}$
and (\ref{eq:su22-c01}) would become 
\begin{equation}
W=2\epsilon _{1}\,(m_{0}m_{1}\Lambda _{1}^{4})^{1/2}-\frac{\mathrm{det}\,(c)\, M_{1}}{m_{0}}-m_{0}M_{0}-m_{1}M_{1}
-\mathrm{tr}\,(c\,T).\label{eq:su22-c01b}
\end{equation}
The result we obtain by minimizing (\ref{eq:su22-c01b}) with  $m_{0}$, $m_{1}$ and  $c$  is again $W=0$ and
exactly  (\ref{eq:su21m-a}).
The lesson is that it does not matter which one chiral superfield we integrate out in computing  $W_{\mathrm{tree,d}}$.
However, we can integrate in the independent gauge singlet matter fields all at one time. 
We will do the same when we consider a general linear 
moose with arbitrary number of nodes later 
in this section and we will give an explicit proof that the final result does not depend on which particular 
chiral superfield we integrate out in computing  $W_{\mathrm{tree,d}}$. 

For Figure \ref{fig-lmoose1}(d), we have $Q_{0}\sim (\square,\, 1)$ and
$Q_{1}\sim (\square,\,\square)$ under the $SU(2)_{1}\times SU(2)_{2}$ gauge
symmetry. The gauge singlets are $M_{0}$, $M_{1}$ and $\mathrm{det}\,(Q_{0}Q_{1})$.
As we will explain later when we discuss a general linear moose with arbitrary number of nodes,
the superpotential can be completely expressed in terms of the gauge singlets $M_{0}$ and $M_{1}$.
The superpotential is then obtained by minimizing
\begin{equation}
W=2\epsilon _{1}\,(m_{0}m_{1}\Lambda _{1}^{4})^{1/2}+2\epsilon _{2}\,(m_{1}\Lambda _{2d}^{5})^{1/2}-m_{0}M_{0}-
m_{1}M_{1}\label{eq:su22-w01}
\end{equation}
with $m_{0}$ and $m_{1}$ which gives
\begin{equation}
W=\frac{\Lambda _{2d}^{5}M_{0}}{M_{0}M_{1}-\Lambda _{1}^{4}}.\label{eq:su22p-a}
\end{equation}
Using the constraint we obtained in (\ref{eq:su21m-a}) for the moduli
space of $SU(2)_{1}$ with two flavors, this can be rewritten as $W=\Lambda _{2d}^{5}M_{0}/
\mathrm{det}\,(Q_{0}Q_{1})$.
Note that (\ref{eq:su22p-a}) contains a single instanton contribution
from $SU(2)_{2}$ and an infinite series of multi-instanton contributions
from $SU(2)_{1}$ as it can be seen by making a Taylor expansion of
$1/(1-\Lambda _{1}^{4}/(M_{0}M_{1}))$ in powers of $\Lambda _{1}^{4}/(M_{0}M_{1})$.

Next let us consider
Figure \ref{fig-lmoose1}(e). The gauge singlets are $M_{0}$,
$M_{1}$, $M_{2}$, $\mathrm{det}\,(Q_{0}Q_{1})$,
$\mathrm{det}\,(Q_{1}Q_{2})$ and the $2\times 2$ matrix
$T=Q_{0}Q_{1}Q_{2}$. As we will explain later, the moduli space constraint we are 
looking for can be parameterized by $M_{0}$,
$M_{1}$, $M_{2}$, and $T$.
Now $T$ is non-quadratic and we can compute
$W_{\mathrm{tree,d}}$ by minimizing
$\mathrm{tr}\,(c\,T)+m_{2}M_{2}$ with $Q_{2}$ which
gives $W_{\mathrm{tree,d}}=-\mathrm{det}\,(c)(M_{0}M_{1}-\Lambda
_{1}^{4})/m_{2}$. 
Integrating out $m_{0}$, $m_{1}$,
$m_{2}$ and $c$ in
\begin{eqnarray}
W & = & 2\epsilon _{1}\,(m_{0}m_{1}\Lambda _{1}^{4})^{1/2}+2\epsilon _{2}\,(m_{1}m_{2}\Lambda _{2}^{4})^{1/2}\nonumber \\
 &  & -\frac{\mathrm{det}\,(c)}{m_{2}}(M_{0}M_{1}-\Lambda _{1}^{4})-m_{0}M_{0}-m_{1}M_{1}
-m_{2}M_{2}-\mathrm{tr}\,(c\,T)\label{eq:su22-mcm2}
\end{eqnarray}
gives $W=0$ and a quantum moduli space constrained by
\begin{equation}
\mathrm{det}\,T-M_{0}M_{1}M_{2}+\Lambda _{1}^{4}M_{2}+\Lambda _{2}^{4}M_{0}=0.\label{eq:su22m-b}
\end{equation}

Finally, let us consider the general case of a linear moose with $r$
nodes and $r+1$ links shown in Figure \ref{fig-lmoose2}.
{\figsize\begin{figure}[htb]
{\centerline{\epsfxsize=15cm \epsfbox{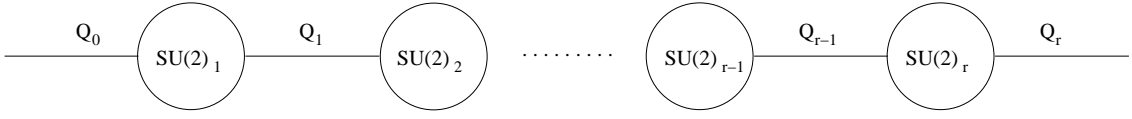}}}
\caption{\figsize\sf \label{fig-lmoose2}
Linear moose with $r$ nodes and $r+1$ links. The external links $Q_0$ and $Q_r$ each have
one color and one subflavor indices and each internal link has
two color indices.
}\end{figure}}
There are $r+1$ link
chiral superfields each with four complex degrees of freedom. We can construct
 a total of $\frac{1}{2}(r^{2}+3r+8)$ gauge singlets given by
determinants of products of one to $r$ consecutive link superfields, and
the product of all the chiral superfields:
\begin{equation}
\mathrm{det }\,(Q_{i}),\label{lin-qi}
\end{equation}
\begin{equation}
\quad\mathrm{det }\,(Q_{i}Q_{i+1}),\quad \,\,\cdots,\,\,
\mathrm{det }\,(Q_{0}Q_{1}\cdots Q_{r-1}),\quad \mathrm{det }\,(Q_{1}Q_{2}\cdots Q_{r}),\label{lin-qis}
\end{equation}
\begin{equation}
\mathrm{ and} \quad
Q_{0}Q_{1}\cdots Q_{r}.\label{eq:linear-s1}\end{equation}

For a generic linear moose, the gauge symmetry is completely
broken and $3r$ of the complex degrees of freedom become massive or are eaten
by the super Higgs mechanism.
Consequently, there are only $4(r+1)-3r=r+4$ massless complex degrees
of freedom left. Because we have $\frac{1}{2}(r^{2}+3r+8)$ gauge
singlets, there must be $\frac{1}{2}(r^{2}+3r+8)-(r+4)=r(r+1)/2$
constraints. We claim that a constraint involving the determinants of only subsegments
of the moose chain given in (\ref{lin-qis}) are not modified by the extra links and nodes.
We can see that as follows: Consider the determinant of a subsegment
$\mathrm{det}\,(Q_{i}Q_{i+1}\cdots Q_{j})$. The color indices from
the gauge groups $SU(2)_{i}$ and $SU(2)_{j}$ are not contracted
with the colors of $SU(2)_{i-1}$ and $SU(2)_{j+1}$ respectively.
Consequently, these adjoining gauge groups behave like global
subflavor symmetries. This amounts to saying that as far as
$\mathrm{det}\,(Q_{i}Q_{i+1}\cdots Q_{j})$ is concerned, the moose
chain is cut off at the $(i-1)^\mathrm{th}$ and $(j+1)^\mathrm{th}$ nodes.
Therefore, finding a constraint for
$\mathrm{det}\,(Q_{i}Q_{i+1}\cdots Q_{j})$ is not an independent
problem. Thus all the $r(r+1)/2$ moduli space constraints can be easily deduced from the one constraint 
which can be parameterized by the $r+5$ 
independent set of gauge singlets:
\begin{equation}
M_{i}\equiv\frac{1}{2}(Q_{i})_{\alpha _{i}\beta _{i}}(Q_{i})_{\alpha' _{i}\beta' _{i}}
\epsilon ^{\alpha _{i}\alpha' _{i}}\epsilon ^{\beta _{i}\beta' _{i}}=\mathrm{det}\,(Q_{i})\label{gaugeinv-1}
\end{equation}
and
\begin{equation}
T_{fg}\equiv \frac{1}{2}(Q_{0})_{f\beta _{0}}(Q_{1})_{\alpha _{1}\beta _{1}}(Q_{2})_{\alpha _{2}\beta _{2}}\cdots (Q_{r})_{\alpha _{r}g}\epsilon ^{\beta _{0}\alpha _{1}}\epsilon ^{\beta _{1}\alpha _{2}}\cdots \epsilon ^{\beta _{r-1}\alpha _{r}}.\label{gaugeinv-2}\end{equation}
where $\alpha_{i}$, $\beta_{i}$ are color indices and $f$, $g$ are subflavor indices.
For $M_{0}$ and $M_{r}$ one of the indices in $Q_{0}$ and $Q_{r}$ is for subflavor.

Now, as we have discussed in detail earlier in this section when we considered the $r=1$ linear moose with two links,  we can  compute $W_{\mathrm{tree,d}}$ by integrating out 
$Q_{k}$ in the gauge and flavor invariant tree
level superpotential
\begin{equation}
\mathrm{tr}\,(c\,T)+m_{k}M_{k},\label{super-nq-1}\end{equation}
where $c$ is a constant $2 \times 2$ matrix and $k$ can take any one value from $0$ to $r$.  This gives \begin{equation}
W_{\mathrm{tree,d}}=-\frac{\mathrm{det}\,(c)}{m_{k}}\,\mathrm{det(}Q_{0}Q_{1}\ldots Q_{k-1})
\,\mathrm{det(}Q_{k+1}Q_{k+2}\ldots Q_{r}).\label{eq:super-nq-2}
\end{equation}
We will see that the final result on the moduli space constraint does
not depend on $k$. Note that (\ref{eq:super-nq-2}) contains two
determinants which can be completely expressed in terms of
$M$'s and $\Lambda$'s.
For simplicity of notation, we introduce a more general way of representing
consecutive products of link chiral superfields and define
\begin{equation}
T_{(i,j)}\equiv Q_{i}Q_{i+1}\ldots Q_{j}.\label{eq:qprod-a}
\end{equation}
Note that $T_{(i,\,j)}$ is a $2 \times 2$ matrix with hidden indices.
The tree level superpotential that contains all the independent gauge invariants is 
\begin{equation}
W_{\mathrm{tree}} = \mathrm{tr}\,(c\,T_{(0,r)})+\sum _{i=0}^{r}m_{i}M_{i}. \label{wtree-all-1}
\end{equation}

The superpotential we need for integrating in all the matter superfields
starting from a pure gauge theory of disconnected nodes is then obtained by using (\ref{sp-1}), (\ref{scales-1}), 
(\ref{eq:super-nq-2}) and (\ref{wtree-all-1}) in (\ref{eq:inter-2a}),
\begin{eqnarray}
 W & =&  2\sum _{i=1}^{r}\epsilon _{i}(\Lambda _{i}^{4}m_{i-1}m_{i})^{\frac{1}{2}}-\frac{\mathrm{det}\,(c)}
{m_{k}}\,\mathrm{det}\,T_{(0,\, k-1)}\,\mathrm{det}\,T_{(k+1,\, r)}\nonumber \\
 &  & -\mathrm{tr}\,(c\,T_{(0,r)})-\sum _{i=0}^{r}m_{i}M_{i}.\label{eq:super-all-1}
\end{eqnarray}
Integrating out $m_{i}$ and $c$ in (\ref{eq:super-all-1}) gives
\[
\epsilon _{1}\,(\frac{\Lambda _{1}^{4}m_{1}}{m_{0}})^{\frac{1}{2}}-M_{0}=0,\]
\[
\epsilon _{1}\,(\frac{\Lambda _{1}^{4}m_{0}}{m_{1}})^{\frac{1}{2}}+\epsilon _{2}\,(\frac{
\Lambda _{2}^{4}m_{2}}{m_{1}})^{\frac{1}{2}}-M_{1}=0,\]
\[
\vdots \]
\[
\epsilon
_{k-1}(\frac{\Lambda_{k-1}^{4}m_{k-2}}{m_{k-1}})^{\frac{1}{2}}+\epsilon
_{k}(\frac{ \Lambda
_{k}^{4}m_{k}}{m_{k-1}})^{\frac{1}{2}}-M_{k-1}=0, \]
\[
\epsilon _{k}(\frac{\Lambda _{k}^{4}m_{k-1}}{m_{k}})^{\frac{1}{2}}+\epsilon _{k+1}
(\frac{\Lambda _{k+1}^{4}m_{k+1}}{m_{k}})^{\frac{1}{2}}+\frac{\mathrm{det}\,(c)}{m_{k}^{2}}
\,\mathrm{det}\,T_{(0,\, k-1)}\,\mathrm{det}\,T_{(k+1,\, r)}-M_{k}=0,\]
\[
\epsilon _{k+1}(\frac{\Lambda _{k+1}^{4}m_{k}}{m_{k+1}})^{\frac{1}{2}}+\epsilon _{k+2}
(\frac{\Lambda _{k+2}^{4}m_{k+2}}{m_{k+1}})^{\frac{1}{2}}-M_{k+1}=0,\]
\[
\vdots \]
\[
\epsilon _{r}(\frac{\Lambda _{r}^{4}m_{r-1}}{m_{r}})^{\frac{1}{2}}-M_{r}=0,\]
\begin{equation}
T_{(0,\, r)}+\frac{\mathrm{det}\,(c)}{m_{k}}c^{-1}\,\mathrm{det}\,T_{(0,\, k-1)}\,\mathrm{det}\,T_{(k+1,\, r)}=0.
\label{eq:minimize-all-a}\end{equation}

Recursively solving for $m_{i}$ and $c$, and putting into (\ref{eq:super-all-1})
gives $W=0$ and a quantum moduli space constrained by
\begin{equation}
\mathrm{det}\,T_{(0,\, r)}-\frac{\mathrm{det}\,T_{(0,\, k-1)}\,\mathrm{det}\,T_{(k+1,\, r)}}{\Omega_{(0,\, k-1)}
\Omega_{(k+1,\, r)}}\Omega_{(0,\, r)}=0,\label{eq:moduli-a}\end{equation}
where we have introduced functions $\Omega_{(i,\, j)}$ to simplify our
notation. The $\Omega$ functions are defined by
\begin{eqnarray}
\Omega_{(i,\, j)} & \equiv  & \prod _{q=i}^{j}M_{q}-\sum _{p=i+1\textrm{ }}^{j}
\Bigl (\Lambda _{p}^{4}\prod _{q\neq p-1,\, p}M_{q}\Bigr )+\sum _{p=i+1\textrm{ }}^{j-2}
\sum _{l=0}^{j-p-2}\Bigl (\Lambda _{p}^{4}\Lambda _{p+l+2}^{4}
\prod _{q\neq p-1,\, p,p+l+1,\, p+l+2}M_{q}\Bigr )\nonumber \\
 &  & -\cdots +(-1)^{(j-i+1)/2}\prod _{p=1}^{(j-i+1)/2}
\Lambda _{i+2p-1}^{4},\textrm{   }\label{eq:om-odd}
\end{eqnarray}
 if $j-i$ is odd, and
\begin{eqnarray}
\Omega_{(i,\, j)} & \equiv  & \prod _{q=i}^{j}M_{q}-\sum _{p=i+1\textrm{ }}^{j}
\Bigl (\Lambda _{p}^{4}\prod _{q\neq p-1,\, p}M_{q}\Bigr )+\sum _{p=i+1\textrm{ }}^{j-2}
\sum _{l=0}^{j-p-2}\Bigl (\Lambda _{p}^{4}\Lambda _{p+l+2}^{4}
\prod _{q\neq p-1,\, p,p+l+1,\, p+l+2}M_{q}\Bigr )\nonumber \\
 &  & -\cdots +(-1)^{(j-i)/2}\sum _{q=0\textrm{ }}^{(j-i)/2}\Bigl (M_{i+2q}
\prod _{p=0}^{q-1}\Lambda _{i+2q-2p-1}^{4}\prod _{l=1}^{(j-i)/2-q}
\Lambda _{i+2q+2l}^{4}\Bigr ),\textrm{  }\label{eq:om-even}
\end{eqnarray}
if $j-i$ is even. We take $j>i$ unless explicitly stated. When $i=j$, we have $\Omega_{(i,\, i)}=\mathrm{det}M_{i}$.
The first few $\Omega $ functions are given in Appendix \ref{sec:ap-om}
and some important recursion relations are given in Appendix \ref{sec:ap-om-prop}.

Thus the quantum moduli space is constrained by the recursion
relations given by (\ref{eq:moduli-a}). Note that $k$ in
(\ref{eq:moduli-a}) is arbitrary and could take any value from $0$
to $r$. As we have argued earlier in this section, a similar relation as
(\ref{eq:moduli-a}) should hold for a
subset of the linear chain, and we write a
more general form of the moduli space constraints as
\begin{equation}
\mathrm{det}\,{T_{(i,j)}}-\frac{\mathrm{det}\,{T_{(i,\, k-1)}}\,\mathrm{det}\,{T_{(k+1,\, j)}}}{
\Omega_{(i,\, k-1)}\Omega_{(k+1,\, j)}}\Omega_{(i,\, j)}=0.\label{eq:moduli-gen}
\end{equation}
Now we can easily prove that the result (\ref{eq:moduli-gen}) is independent
of $k$, since we can repeatedly use the same recursion relations to simplify the fractional factor in the
second term, 
and (\ref{eq:moduli-gen}) gives
\begin{equation}
\mathrm{det}\,T_{(i,j)}-\Omega_{(i,\,j)}=0.\label{eq:moduli-simple}
\end{equation}
Note that (\ref{eq:moduli-simple}) gives $r(r+1)/2$
constraints that completely remove all the redundancy in the set
of gauge singlets.


\setcounter{equation}{0}
\section{\label{sec:intout-links} Integrating out link fields}

In this section we will see that the quantum moduli space constraints of the
linear moose we found in Section \ref{sec-linear-su2r}  give correct and known dynamical
superpotentials when we integrate out some link chiral
superfields. We will consider only the
cases of $r=1$ and $r=2$, since we can compare the results with established dynamical
superpotentials in these cases. The low energy superpotentials we will obtain
after integrating out the link fields are correct and consistent with the results in
Section \ref{sec-linear-su2r} and \cite{ILS-1}.
We will integrate out the external links
in a linear moose with arbitrary number of nodes in Section \ref{sec:ring}.

First let us consider 
the $r=1$ linear moose shown in Figure \ref{fig-lmoose1}(c). This theory
has a quantum moduli space of vacua given by (\ref{eq:su21m-a}).
The superpotential can be written as
\begin{equation}
W=A(\mathrm{det}\,T_{(0,\,1)}-M_{0}M_{1}+\Lambda _{1}^{4}),\label{su21m-a2}
\end{equation}
where $A$ is a Lagrange multiplier. Integrating out $A$ in (\ref{su21m-a2}) simply
gives the constraint (\ref{eq:su21m-a}).
We integrate out $Q_{1}$ by minimizing
\begin{equation}
W=A(\mathrm{det}\,T_{(0,\,1)}-M_{0}M_{1}+\Lambda _{1}^{4})+m_{1}M_{1}
\label{eq:sintout-1}\end{equation}
with $M_{1}$, $T_{(0,\,1)}$ and $A$ to obtain
\begin{equation}
-A\,M_{0}+m_{1}=0,\quad A\,T_{(0,\,1)}^{-1}\,\mathrm{det}\,T_{(0,\,1)}=0,\quad
\mathrm{det}\,T_{(0,\,1)}-M_{0}M_{1}+\Lambda _{1}^{4}=0\label{eq:sintout-2}\end{equation}
with solution \begin{equation}
M_{1}=\frac{\Lambda _{1}^{4}}{M_{0}},\quad
\mathrm{det}\,T_{(0,\,1)}=0,\quad A=\frac{m_{1}}{M_{0}}.\label{eq:sintout-3}
\end{equation}
Putting (\ref{eq:sintout-3}) in (\ref{eq:sintout-1}) gives exactly
the superpotential of a single node with one link given in (\ref{eq:su21up}).

If we choose to integrate
out both $Q_{0}$ and $Q_{1}$ at the same time, we minimize
\begin{equation}
W=A(\mathrm{det}\,T_{(0,\,1)}-M_{0}M_{1}+\Lambda _{1}^{4})
+m_{0}M_{0}+m_{1}M_{1}
\label{eq:sintout-4}\end{equation}
with $M_{0}$, $M_{1}$, $T_{(0,\,1)}$ and $A$ to obtain the same equations
as in (\ref{eq:sintout-2}) and
\begin{equation}
-A\,M_{1}+ m_{0}=0.\label{eq:sintout-5}
\end{equation}
There are two sets of solutions given by 
\begin{equation}
M_{0}=\epsilon\,(\frac{m_{1}\Lambda _{1}^{4}}{m_{0}})^{1/2},\quad
M_{1}=\epsilon\,(\frac{m_{0}\Lambda _{1}^{4}}{m_{1}})^{1/2},\quad
\mathrm{det}\,T_{(0,\,1)}=0,\quad A=\epsilon\,(\frac{m_{0}m_{1}}
{\Lambda _{1}^{4}})^{1/2},\label{eq:sintout-6}\end{equation}
where $\epsilon=\pm1$. Putting (\ref{eq:sintout-6}) in (\ref{eq:sintout-4}) gives exactly
(\ref{sp-1}) for $r=1$.

Next let us consider
the $r=2$ linear moose with  external links shown in Figure \ref{fig-lmoose1}(e).
First let us integrate out $Q_{2}$. This is done by minimizing
\begin{equation}
W=A(\mathrm{det}\,T_{(0,\,2)}-M_{0}M_{1}M_{2}+\Lambda _{1}^{4}M_{2}+\Lambda _{2}^{4}M_{0})
+m_{2}M_{2}
\label{eq:super-q2-1}
\end{equation}
with $M_{2}$, $T_{(0,\,2)}$ and $A$ which gives exactly (\ref{eq:su22p-a}).
Note also that the superpotential (\ref{eq:su22p-a}) vanishes if
we set $\Lambda _{2d}\equiv 0$; and the theory with one node of $SU(2)_{1}$ linked to $Q_{0}$
and $Q_{1}$ has a quantum moduli space as expected and seen in (\ref{eq:su21m-a}).
On the other hand, if we set $\Lambda _{1}\equiv 0$ in (\ref{eq:su22p-a}),
we obtain $W=\Lambda _{2d}^{5}/M_{1}$ which is exactly the superpotential
of $SU(2)_{2}$ with a single flavor. We can further integrate out
$Q_{0}$ and obtain a moose diagram with two nodes and an internal
link. This is done by minimizing
\begin{equation}
\frac{\Lambda _{2d}^{5}M_{0}}{M_{0}M_{1}-\Lambda _{1}^{4}}+m_{0}M_{0}\label{eq:super-q2-add1}\end{equation}
with $M_{0}$ which gives \begin{equation}
W=\frac{\Lambda _{1d}^{5}}{M_{1}}+\frac{\Lambda _{2d}^{5}}{M_{1}}\pm
2\frac{(\Lambda _{1d}^{5}\Lambda _{2d}^{5})^{1/2}}{M_{1}},\label{eq:wintl1l2}
\end{equation}
where $\Lambda _{1d}^{5}=\Lambda _{1}^{4}m_{0}$. Note that in this
case the original $SU(2)_{1}\times SU(2)_{2}$ gauge symmetry is
broken by $M_{1}$ into a diagonal $SU(2)_{D}$. The first term in
(\ref{eq:wintl1l2}) comes from a single instanton contribution in
the completely broken $SU(2)_{1}$, the second term also comes from
a single instanton in the completely broken $SU(2)_{2}$, and the
last term comes from gaugino condensation in the unbroken
$SU(2)_{D}$. By threshold matching the gauge couplings at energy
$M_{1}^{1/2}$,
${\Lambda{_D}}^{6}/M_{1}^{3}=e^{-8{\pi}^{2}/{g_{D}^{2}}}
=e^{-8{\pi}^{2}(g_{1}^{-2}+g_{2}^{-2})}={{\Lambda_{1d}}^5
{\Lambda_{2d}}^5/M_1^{5}}$, where we used
$g_{D}^{-2}=g_{1}^{-2}+g_{2}^{-2}$ for the gauge coupling constants, we see that 
the scale of the
low energy $SU(2)_{D}$ is $\Lambda _{D}=[(\Lambda _{1d}^{5}\Lambda
_{2d}^{5})^{1/2}/M_{1}]^{1/3}$.

We can integrate out $Q_{1}$ instead of $Q_{2}$ by minimizing
\begin{equation}
W=A(\mathrm{det}\,T_{(0,\,2)}-M_{0}M_{1}M_{2}+\Lambda _{1}^{4}M_{2}+\Lambda _{2}^{4}M_{0})+
m_{1}M_{1}\label{eq:superp-q1-1}
\end{equation}
with $M_{1},$ $T_{(0,\,2)}$ and $A$, which gives
\begin{equation}
W=\frac{\Lambda _{1d}^{5}}{M_{0}}+\frac{\Lambda _{2d}^{5}}{M_{2}},\label{eq:superp-q1-2}\end{equation}
where $\Lambda _{id}=(\Lambda _{i}^{4}m_{1})^{1/5}$
are the scales for the low energy theory. (\ref{eq:superp-q1-2})
is exactly the superpotential for two disconnected gauge groups with
a single flavor attached to each. We can further integrate
out the two remaining fields $M_{0}$ and $M_{2}$ by adding $m_{0}M_{0}+m_{2}M_{2}$
to (\ref{eq:superp-q1-2}) and minimizing with $M_{0}$ and $M_{2}$.
This gives
\begin{eqnarray}
W & = & 2\epsilon _{1}\Lambda _{01}^{3}+2\epsilon _{2}\Lambda _{02}^{3},\label{eq:superp-gc-2}
\end{eqnarray}
where $\Lambda _{01}=(\Lambda _{1d}^{5}m_{0})^{1/6}$ and $\Lambda _{02}=(\Lambda _{2d}^{5}m_{2})^{1/6}$
are the scales of the pure $SU(2)_{1}$ and $SU(2)_{2}$ gauge theories
respectively.

We can, if we wish, integrate out all the matter fields at the same time
by minimizing the superpotential
\begin{equation}
W  =  A(\mathrm{det}\,T_{(0,\,2)}-M_{0}M_{1}M_{2}+\Lambda _{1}^{4}M_{2}+\Lambda _{2}^{4}M_{0})
 +m_{0}M_{0}+m_{1}M_{1}+m_{2}M_{2}
\label{eq:superp-quant-2b}
\end{equation}
with $M_{0}$, $M_{1}$, $M_{2}$, $T_{(0,\,2)}$ and $A$. We obtain four sets of solutions
\[
M_{0} =\epsilon _{1} (\frac{\Lambda _{1}^{4}m_{1}}{m_{0}})^{\frac{1}{2}}, \quad 
M_{2}=\epsilon _{2}\,(\frac{\Lambda _{2}^{4}m_{1}}{m_{2}})^{\frac{1}{2}},\mathrm{  }\quad\]
\begin{equation}
M_{1}=\epsilon _{1}\,(\frac{\Lambda _{1}^{4}m_{0}}{m_{1}})^{\frac{1}{2}}+\epsilon _{2}\,(\frac{
\Lambda _{2}^{4}m_{2}}{m_{1}})^{\frac{1}{2}},\quad
\mathrm{det}\,T_{(0,\,2)} =0, \quad A=\epsilon _{1}\epsilon _{2}\,(\frac{m_{0}m_{2}}{\Lambda _{1}^{4}
\Lambda _{2}^{4}})^{\frac{1}{2}}.
\label{eq:vevm2-1a}
\end{equation}
Putting (\ref{eq:vevm2-1a}) in (\ref{eq:superp-quant-2b})
gives exactly (\ref{eq:superp-gc-2}) with $\Lambda _{01}=(\Lambda _{1}^{4}m_{0}m_{1})^{1/6}$
and $\Lambda _{02}=(\Lambda _{2}^{4}m_{1}m_{2})^{1/6}$.
Thus we have consistently reproduced the pure low energy dynamical
superpotential.


\setcounter{equation}{0}
\section{\label{sec:perturbations}Tree level perturbations }

In this section we will study the vacuum structure of the $r=2$ linear moose shown in Figure \ref{fig-lmoose1}(e)
when perturbed by the tree level superpotential
\begin{equation}
W_{\mathrm{tree}}=m_{0}M_{0}+m_{1}M_{1}+m_{2}M_{2}+\mathrm{tr}\,(c\,T_{(0,2)}),\label{eq:superp-tree-1}
\end{equation}
which includes a non-zero coupling to the non-quadratic gauge singlet $T_{(0,2)}$. The lesson we will learn is that
the inclusion of the non-quadratic gauge singlet term in $W_{\mathrm{tree}}$ leads to discrete vacua and also the 
math becomes complicated. We will explicitly compute the discrete vacua. 

Semi-classically, there are two vacuum states. One is at the origin,
\begin{equation}
M_{0}=M_{1}=M_{2}=T_{(0,2)}=0,\label{eq:pert-classmt0}\end{equation}
 where the original $SU(2)_{1}\times SU(2)_{2}$ gauge symmetry is
preserved. The second vacuum is at
\begin{equation}
M_{0}=\frac{m_{1}m_{2}}{\mathrm{det}\,(c)},\quad M_{1}=\frac{m_{0}m_{2}}{\mathrm{det}\,(c)},
\quad M_{2}=\frac{m_{0}m_{1}}{\mathrm{det}\,(c)},\quad T_{(0,2)}=\frac{m_{0}m_{1}m_{2}}{\mathrm{det}\,(c)}c^{-1}
\label{eq:pert-classmt}
\end{equation}
where the gauge symmetry is completely broken.

In the quantum theory,
the vacuum structure is much richer. The vacuum expectation values
in the quantum theory perturbed by (\ref{eq:superp-tree-1}) are obtained
by minimizing the superpotential
\begin{equation}
W=A(\mathrm{det}\,T_{(0,2)}-M_{0}M_{1}M_{2}+\Lambda _{1}^{4}M_{2}+\Lambda _{2}^{4}M_{0})
+m_{0}M_{0}+m_{1}M_{1}+m_{2}M_{2}+\mathrm{tr}\,(c\,T_{(0,2)})\label{eq:superp-quant-2}
\end{equation}
with $M_{0}$, $M_{1}$, $M_{2}$, $T_{(0,2)}$ and $A$. The solution is given in Appendix \ref{sec:ap-pert}. All we
need for our discussion here is that the expectation values of $M_{0}$, $M_{1}$, $M_{2}$, $T_{(0,2)}$ and $A$ can
be parameterized by $x$ such that \begin{eqnarray}
 &  & \Lambda _{1}^{8}\Lambda _{2}^{8}m_{1}x^{5}-2\Lambda _{1}^{4}\Lambda _{2}^{4}m_{0}m_{1}m_{2}x^{3}
-\Lambda _{1}^{4}\Lambda _{2}^{4}\textrm{det}\,(c)^{2}x^{2}\nonumber \\
 &  & -(\Lambda _{1}^{4}m_{0}\textrm{det}\,(c)^{2}+\Lambda _{2}^{4}m_{2}\textrm{det}\,(c)^{2}
-m_{0}^{2}m_{1}m_{2}^{2})x-m_{0}m_{2}\textrm{det}\,(c)^{2}=0.\label{eq:x-1}
\end{eqnarray}

Note how messy the solution given in Appendix \ref{sec:ap-pert} is even for the case of only
two nodes. The complication comes because of the presence of $\mathrm{tr}\,(c\,T_{(0,2)})$ in $W_{\mathrm{tree}}$.
There are in general five solutions to $x$ which give five sets of
expectation values with non-zero $\mathrm{det}\,(T_{(0,\,2)})$ and the vacuum space becomes discrete in the perturbed
theory.
Let us simplify and interpret the expectation
values in some limits of the coupling constants.
If the mass $m_{1}$ is set to zero, there are only two solutions
to (\ref{eq:x-1}) given by
\begin{equation}
x=(-\frac{m_{0}}{\Lambda _{2}^{4}},\, -\frac{m_{2}}{\Lambda _{1}^{4}})\label{eq:pert-case1x}\end{equation}
 which give, using Appendix \ref{sec:ap-pert}, two vacua at
\[
 M_{0}=(-\frac{\mathrm{det}\,(c)\Lambda _{2}^{4}}{m_{0}^{2}},\, 0),\quad M_{1}=
(\frac{m_{0}m_{2}-m_{0}^{2}\Lambda _{1}^{4}/\Lambda _{2}^{4}}{\,\mathrm{det}\,(c)},\,
\frac{m_{0}m_{2}-m_{2}^{2}\Lambda _{2}^{4}/\Lambda _{1}^{4}}{\,\mathrm{det}\,(c)}),\]
\begin{equation}
 M_{2}= (0,\,-\frac{\mathrm{det}\,(c)\Lambda _{1}^{4}}{m_{2}^{2}}),\quad
T_{(0,2)}=(\frac{\mathrm{det}\,(c)\Lambda _{2}^{4}}{m_{0}}c^{-1},\, \frac{\mathrm{det}\,(c)
\Lambda _{1}^{4}}{m_{2}}c^{-1}).
\label{eq:pert-case1mt}
\end{equation}
Note that for large $m_{0}$ and $m_{2}$ in (\ref{eq:pert-case1mt}), the expectation values of $M_{0}$, $M_{2}$ and
$T_{(0,2)}$ vanish and the links $Q_{0}$ and $Q_{2}$ are missing in the low energy theory.
In this case, we have only  the internal link $Q_{1}$. The
$SU(2)_{1}\times SU(2)_{2}$ gauge symmetry is  then broken by $M_{1}$ down to a diagonal
$SU(2)_{D}$. Gaugino condensation in $SU(2)_{D}$ breaks the $Z_{4}$ $R$
symmetry to $Z_{2}$. 

If we set $m_{0}=m_{2}=0$ in (\ref{eq:x-1}), then 
\begin{equation}
x=(0,\, 0,\, (\frac{\textrm{det}c^{2}}
{m_{1}\Lambda _{1}^{4}\Lambda _{2}^{4}})^{\frac{1}{3}},\, e^{i\pi /3}(\frac{\textrm{det}c^{2}}
{m_{1}\Lambda _{1}^{4}\Lambda _{2}^{4}})^{\frac{1}{3}},\, e^{i2\pi /3}(\frac{\textrm{det}c^{2}}
{m_{1}\Lambda _{1}^{4}\Lambda _{2}^{4}})^{\frac{1}{3}})
\label{eq:pert-casem0m2}
\end{equation}
and there are four distinct vacua.
In this case, $SU(2)_{1}\times SU(2)_{2}$ gauge symmetry is completely broken.
For large $m_{1}$,  we have $M_{1}=0$ and there are two disconnected $SU(2)$'s with a single
link attached to each in the low energy theory.

If we set $\mathrm{det}\,(c) = 0$ in (\ref{eq:x-1}), we obtain 
\begin{equation}
x=(0,\, -(\frac{m_{0}m_{2}}{\Lambda _{1}^{4}\Lambda _{2}^{4}})^{\frac{1}{2}},\, -(\frac{m_{0}m_{2}}{
\Lambda _{1}^{4}\Lambda _{2}^{4}})^{\frac{1}{2}},\, (\frac{m_{0}m_{2}}{\Lambda _{1}^{4}\Lambda _{2}^{4}})^{\frac{1}{2}},\,
(\frac{m_{0}m_{2}}{\Lambda _{1}^{4}\Lambda _{2}^{4}})^{\frac{1}{2}}).\label{eq:pert-case3x}
\end{equation}
In this case, all the links are absent in the low energy theory and 
the $Z_{4} \times Z_{4}$ $R$ symmetry is broken 
down to $Z_{2} \times Z_{2}$  by gaugino condensation in the two nodes. 
The vacua at $x=\pm (m_{0}m_{2}/{(
\Lambda _{1}^{4}\Lambda _{2}^{4}))^{1/2}}$ are near the origin and they correspond to the semi-classical
vacuum  at the origin. 
The vacuum state at $x=0$ is far out in moduli space and it corresponds to the second semi-classical vacuum.


\setcounter{equation}{0}
\section{\label{sec:ring}Ring moose}

Now we can construct the quantum  moduli space of the ring moose shown in Figure \ref{figmring} starting from
the linear moose shown in Figure \ref{fig-lmoose2}. First we list $r^2+1$ 
gauge singlets in the ring moose given by $M_{i}$ defined in (\ref{gaugeinv-1}), where $0 \le i \le r-1$ now,
\begin{equation}
U_{(0,\,r-1)}\equiv
 \frac{1}{2}(Q_{0})_{\alpha _{0}\beta _{0}}(Q_{1})_{\alpha _{1}\beta _{1}}
(Q_{2})_{\alpha _{2}\beta _{2}}\cdots (Q_{r-1})_{\alpha _{r-1}\beta _{r-1}}
\epsilon ^{\beta _{0}\alpha _{1}}\epsilon ^{\beta _{1}\alpha _{2}}\cdots
\epsilon ^{\beta _{r-1}\alpha _{0}}, \label{eq:eq:u0r}\end{equation}
and
\begin{equation}
\mathrm{det }\,(Q_{i}Q_{i+1}),\quad \mathrm{det }\,(Q_{i}Q_{i+1}Q_{i+2}),\,\,\cdots,\,\,
\mathrm{det }\,(Q_{i}Q_{i+1}\cdots Q_{i-1}).\label{eq:ring-s1}\end{equation}
We identify $i\sim i+r$.
We have already found the constraints that relate the determinants listed in (\ref{eq:ring-s1})
to $M_{i}$ and $\Lambda_{j}$
in Section \ref{sec-linear-su2r}.
Therefore, we only need to find the one
constraint that relates $U_{(0,\,r-1)}$
to $M_i$ and $\Lambda_{j}$. In fact, one can check that there are only $r+1$ independent gauge invariant as follows:
The link chiral superfields have a total of $4r$ complex components. On the other hand, there are 
$3r$ D-flatness conditions and only $3r-1$ of these conditions are independent because of the unbroken $U(1)$ 
gauge symmetry. Thus there must be $4r-(3r-1)=r+1$ independent complex degrees of freedom which we can 
choose as $U_{(0,\,r-1)}$ and $M_i$. 

We will start with the moduli space of the linear moose with
external links found in Section \ref{sec-linear-su2r}. We will then
integrate out the external links and construct the superpotential
for the moose with only internal links shown in
Figure \ref{figmringa}(b). Finally, a link field
that transforms as $(\square,\,\square)$ under $SU(2)_{r}\times
SU(2)_{1}$ will be integrated in to build the ring moose shown in Figure \ref{figmring}.
Since we can at the same time compute the superpotential with only
one external link, let us first integrate out $Q_{r}$ and obtain
the superpotential for Figure \ref{figmringa}(a).
{\figsize\begin{figure}[htb]
{\centerline{\epsfxsize=13cm \epsfbox{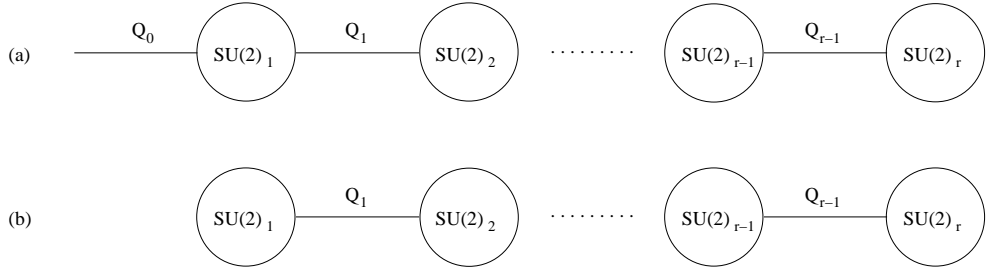}}}
\caption{\figsize\sf\label{figmringa}  (a) Linear moose with $r$ nodes and one external link.
(b) Linear moose with only internal links.  The external link $Q_0$ has
one color and one subflavor indices and each internal link has
two color indices.
}\end{figure}}
This is done by integrating out $M_{r}$, $T_{(0,\,r)}$ and $A$ in
\begin{equation}
W=A\Bigl (\mathrm{det}\,{T_{(0,\, r)}}-\Omega_{(0,\,r)}\Bigr
)+m_{r}M_{r}.\label{eq:winternalmin}
\end{equation}
As shown in Appendix \ref{sec:ap-wmt}, the resulting superpotential is
\begin{equation}
W=\frac{\Lambda _{rd}^{5}\Omega_{(0,\, r-2)}}{\Omega_{(0,\, r-1)}},\label{eq:woneext}
\end{equation}
where $\Lambda _{rd}^{5}=\Lambda _{r}^{4}m_{r}$.

Next we integrate
out $Q_{0}$ by adding $m_{0}M_{0}$ to (\ref{eq:woneext}) and minimizing
with $M_{0}$ which, as shown in Appendix \ref{sec:ap-wm}, gives
the superpotential for Figure \ref{figmringa}(b),
\begin{equation}
W=\frac{\Lambda _{1d}^{5}\Omega_{(2,\, r-1)}}{\Omega_{(1,\, r-1)}}+\frac{\Lambda _{rd}^{5}
\Omega_{(1,\, r-2)}}{\Omega_{(1,\, r-1)}}\pm 2\frac{(\Lambda _{1d}^{5}\Lambda _{rd}^{5}
\prod _{i=2}^{r-1}\Lambda _{i}^{4})^{1/2}}{\Omega_{(1,\, r-1)}},\label{eq:winternal}
\end{equation}
where $\Lambda _{1d}^{5}=\Lambda _{1}^{4}m_{0}$. 
This superpotential  can be interpreted
as follows: For the moose chain
shown in Figure \ref{figmringa}(b), the original $SU(2)^{r}$ gauge symmetry is completely broken and there is 
a new unbroken diagonal
$SU(2)_{D}$.  The first and second terms come from a single instanton in the broken
$SU(2)_{1}$ and  a single instanton in the broken $SU(2)_{r}$ respectively and infinite series of 
multi-instantons from the broken
$SU(2)_{2}$ to  $SU(2)_{r-2}$. 
This can be seen by using the explicit form of the
$\Omega$ functions and making an expansion of $\Omega_{(1,\, r-1)}^{-1}$ in powers of the
scales of $SU(2)_{2}$ to  $SU(2)_{r-2}$ .
 The last term comes from gaugino condensation
in the unbroken diagonal $SU(2)_{D}$. In fact, we can read off from  (\ref{eq:winternal}) that
the scale of the diagonal $SU(2)_{D}$ is
\begin{equation}
\Lambda_{D}=\Bigl(\frac{(\Lambda _{1d}^{5}\Lambda _{rd}^{5}
\prod _{i=2}^{r-1}\Lambda _{i}^{4})^{1/2}}{\Omega_{(1,\, r-1)}}\Bigr)^{\frac{1}{3}}.\label{lambda-all-1}
\end{equation}

Finally, we can construct the quantum moduli space of the ring moose
 shown in Figure \ref{figmring}.
{\figsize\begin{figure}[htb]
{\centerline{\epsfxsize=6cm
\epsfbox{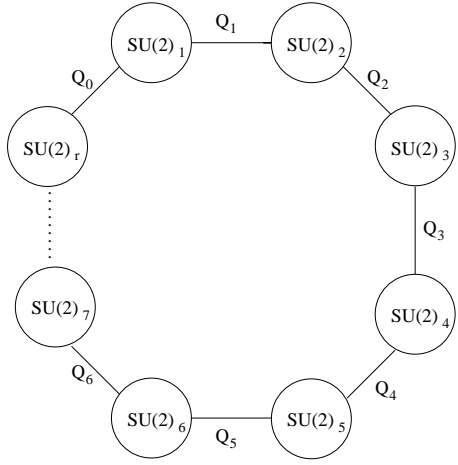}}} \caption{\figsize\sf\label{figmring} Ring
moose with $r$ nodes.  Each link has two color indices. }\end{figure}}
The tree level superpotential that contains the new gauge invariant fields is
\begin{equation}
W_{\mathrm{tree }}=b\,U_{(0,\,r-1)}+m_{0}M_{0}.\label{ring-tree-1a}
\end{equation}
where $b$ and $m_{0}$  are  constants.
Because $U_{(0,\,r-1)}$ is a non-quadratic singlet, minimizing
$b\,U_{(0,\,r-1)}+m_{0}M_{0}$ with $Q_{0}$ gives
\begin{equation}
W_{\mathrm{tree,d }}=-\frac{b^{2}}{4m_{0}}\Omega_{(1,\, r-1)}.\label{wtreedring}
\end{equation}
The superpotential we need to integrate in $Q_{0}$ is then obtained by
setting $\Lambda _{1d}^{5} \rightarrow m_{0}\Lambda _{1}^{4}$,  
$\Lambda _{rd}^{5} \rightarrow m_{0}\Lambda _{r}^{4}$ in (\ref{eq:winternal}), adding 
(\ref{wtreedring}) and subtracting (\ref{ring-tree-1a}),
\begin{equation}
W = \frac{m_{0}\Lambda _{1}^{4}\Omega_{(2,\, r-1)}}{\Omega_{(1,\, r-1)}}+
\frac{m_{0}\Lambda _{r}^{4}\Omega_{(1,\, r-2)}}{\Omega_{(1,\, r-1)}}\pm 2\frac{m_{0}(
\prod _{i=1}^{r}\Lambda _{i}^{4})^{1/2}}{\Omega_{(1,\, r-1)}} 
-\frac{b^{2}}{4m_{0}}\Omega_{(1,\, r-1)}-m_{0}M_{0}-b\,U_{(0,\,r-1)}.\label{eq:wminring}
\end{equation}
Minimizing (\ref{eq:wminring}) with $m_{0}$ and $b$ gives, see Appendix \ref{sec:ap-ring},
$W=0$ and \begin{equation}
U_{(0,\,r-1)}^{2}+\Lambda _{1}^{4}\Omega_{(2,\, r-1)}+\Lambda _{r}^{4}\Omega_{(1,\, r-2)}-M_{0}
\Omega_{(1,\, r-1)}\pm 2(\prod _{i=1}^{r}\Lambda _{i}^{4})^{1/2}=0.\label{eq:moduli-ring-b}
\end{equation}
This is symmetric in all links and scales.

Before we interpret (\ref{eq:moduli-ring-b}), let us first recall that
according to the
Seiberg-Witten hypothesis \cite{SW-1}, the quantum moduli space of
an $SU(2)$ gauge theory with $\mathcal{N}=2$ supersymmetry coincides with the moduli space of the
elliptic curve $y^{2}=(x^{2}-u)^{2}-\Lambda ^{4}$,
where $u$ is a gauge invariant coordinate and $\Lambda$ is the 
dynamical scale of the theory. 
The singularities of this curve are given by the zeros of the
discriminant $\Delta _{\Lambda }=(u^{2}-\Lambda ^{4})(2\Lambda)
^{8}$. This occurs at $u=\pm \Lambda ^{2}$ and $u=\infty $. The
first two singularities at $u=\pm \Lambda ^{2}$ are in the strong
coupling region, and there is a massless monopole at one and a
massless dyon at the other of these singularities. The singularity
at $u=\infty $ is in the semi-classical region. 
 
Now let us rewrite (\ref{eq:moduli-ring-b}) as
\begin{equation}
u_{r} = \pm \Lambda_{(1,\,r)}^{2},\label{eq:u-lam-1}
\end{equation}
where
\begin{equation}
u_{r} \equiv U_{(0,\,r-1)}^{2}+\Lambda _{1}^{4}\Omega_{(2,\, r-1)}+\Lambda
_{r}^{4}\Omega_{(1,\, r-2)}-M_{0} \Omega_{(1,\, r-1)}\quad \mathrm{and}\quad 
\Lambda_{(1,\,r)}^{2} \equiv 2(\prod _{i=1}^{r}\Lambda _{i}^{4})^{1/2}.
\label{eq:ufun-lam}
\end{equation}

Note that the modulus $u_{r}$ contains all the independent gauge invariants we needed to parameterize
the moduli space of the ring. What (\ref{eq:u-lam-1}) is telling us is that the
function $u_{r}$ is locked at $\pm \Lambda_{(1,\,r)}^{2}$.
In other words, (\ref{eq:u-lam-1}) gives two $r$ - complex dimensional singular submanifolds in the $r+1$ - complex 
dimensional moduli space spanned by all the independent gauge invariants. 
The vacua are fixed to the singularities because of the tree level deformation of the theory.  
That is why the integrating in
procedure is relevant to the Seiberg-Witten curve.%
\footnote{I like to thank the referee for suggesting the last two statements.}%
Giving large VEVs to the link fields breaks the original $SU(2)^{r}$
gauge symmetry into a diagonal $SU(2)_{D}$ with matter in the adjoint representation. 
The two singularities given by (\ref{eq:u-lam-1}) on the $u_{r}$ plane 
can be nothing but the two singularities in the strong coupling region of the $SU(2)_{D}$ gauge theory with 
$\mathcal{N}=2$ supersymmetry. 
The monodromies around these singularities on the $u_{r}$ plane must be the same as in 
Seiberg-Witten and 
the charge at
the singularity  $u_{r} = + \Lambda_{(1,\,r)}^{2}$
is that of a monopole and the charge at $u_{r} = - \Lambda_{(1,\,r)}^{2}$ is that of a dyon. 
A generic point in the moduli space of 
the ring moose has unbroken $U(1)$ gauge symmetry and the ring moose is 
in the Coulomb phase.
Having obtained these singularities and because the $U(1)$ coupling coefficient is holomorphic, we have 
determined the 
elliptic curve that
parameterizes the Coulomb phase of the ring moose.  

Thus the quantum moduli space of the
ring moose can be parameterized by the elliptic curve
\begin{equation}
y^{2}=\Bigl (x^{2}-[U_{(0,\,r-1)}^{2}+\Lambda _{1}^{4}\Omega_{(2,\, r-1)}+\Lambda _{r}^{4}
\Omega_{(1,\, r-2)}-M_{0}\Omega_{(1,\, r-1)}]\Bigr )^{2}-4\prod _{i=1}^{r}\Lambda _{i}^{4}.
\label{eq:ssw-curve-a}\end{equation}
The first few $u$
functions are listed in Appendix \ref{sec:ap-u}.

Seiberg-Witten curves for the ring moose were computed in \cite{CEFS} using a different method. 
A method used in \cite{IS-1} to obtain the curve for the $r=2$ ring was continued in \cite{CEFS} 
to compute the curve for $r=3$. 
The idea was as follows:
Because giving large VEVs to the link fields breaks the
$SU(2)^{r}$ gauge symmetry into a diagonal
$SU(2)_{D}$ with matter in the adjoint representation, the theory
in effect becomes that of a single $SU(2)$ with $\mathcal{N}=2$
supersymmetry. The curve for $r=3$ was obtained by taking various asymptotic limits of 
the gauge singlet fields and the 
nonperturbative scales, comparing with
the $\mathcal{N}=2$ $SU(2)$ curve and imposing symmetries. 
The result for $r=3$ was then generalized
to the curve for a ring moose with arbitrary $r$.
Our results agree  with
\cite{IS-1} for $r=2$ and with \cite{CEFS} for $r=3$. However, we
do not agree with the curves in \cite{CEFS} for $r\geq 4$. Only few
terms in $u_{r}$ were obtained in \cite{CEFS}, which would give incorrect singular submanifolds in
moduli space. We are not suggesting that the method used in \cite{CEFS} is incorrect.
Furthermore, work on applications to deconstruction \cite{CEGK} - \cite{CEKPSS} should not be affected by
the missing terms as they did not rely on the 
parameterization of the modulus in terms of the independent gauge invariant coordinates.
Here we have obtained the quantum moduli space directly by integrating in all
the independent link fields starting from a pure gauge theory of disconnected nodes and building the
ring moose via the linear moose. This is done for a ring with arbitrary
number of nodes without any need of imposing symmetries in the
nodes or links and without taking asymptotic limits; and the result is
automatically symmetric in all nodes and links.


\setcounter{equation}{0}
\section{Monopole condensates\label{moncon}}

We will now look at the effective field theory near the singularities
of the quantum moduli space of the ring moose. It is believed that the singularities
correspond to massless particles \cite{SW-1} and
there is a massless monopole at 
$u_{r} = + \Lambda_{(1,\,r)}^{2}$
and a massless 
dyon at $u_{r} = - \Lambda_{(1,\,r)}^{2}$.  As these massless states move out of the singularities, 
they get masses of order
 $u_{r} \mp \Lambda_{(1,\,r)}^{2}$. 
The leading order effective superpotential
near the singularities can thus be written as
\begin{eqnarray}
W & \sim & \Bigl (U_{(0,\, r-1)}^{2}+\Lambda _{1}^{4}\Omega_{(2,\, r-1)}+\Lambda _{r}^{4}
\Omega_{(1,\, r-2)}-M_{0}\Omega_{(1,\, r-1)}- \Lambda _{(1,\, r)}^{2}\Bigr )
\tilde{E}_{\mathrm{m} }E_{\mathrm{m} }  \nonumber\\
& & + \Bigl (U_{(0,\, r-1)}^{2}+\Lambda _{1}^{4}\Omega_{(2,\, r-1)}+\Lambda _{r}^{4}
\Omega_{(1,\, r-2)}-M_{0}\Omega_{(1,\, r-1)}+ \Lambda _{(1,\, r)}^{2}\Bigr )
\tilde{E}_{\mathrm{d} }E_{\mathrm{d} }+\sum _{i=0}^{r-1}m_{i}M_{i}, \label{eq:moncon-1}
\end{eqnarray}
where $E_{\mathrm{m}}$ and $E_{\mathrm{d}}$ are chiral superfields of the monopole and dyon states
respectively. The last (mass) term in (\ref{eq:moncon-1}) is added to lift up the flat directions
and give nonzero VEV to the condensates. 
Note that although the singularities look the same as in Seiberg-Witten on the $u_{r}$ plane, they are 
$r$ - complex dimensional submanifolds with a very nontrivial modulus given by (\ref{eq:ufun-lam}).   
The equations of motion are obtained by minimizing (\ref{eq:moncon-1})
with $M_{i}$, $U_{(0,\, r-1)}$,  $E_{\mathrm{m}}$, $\tilde{E}_{\mathrm{m}}$,
$E_{\mathrm{d}}$ and $\tilde{E}_{\mathrm{d}}$ which,
using the  properties of the $\Omega$ functions given in Appendix \ref{sec:ap-om-prop},
gives two sets of equations. One for the first singularity at  $u_{r} = + \Lambda_{(1,\,r)}^{2}$ with $E_{\mathrm{d}}=0$ and
\[-\Omega_{(0,\, i-1)}\Omega_{(i+1,\, r-1)}\tilde{E}_{\mathrm{m} }E_{\mathrm{m} }+m_{i}=0,
\]
\[
U_{(0,\, r-1)}=0,\]
\begin{equation}
U_{(0,\, r-1)}^{2}+\Lambda _{1}^{4}\Omega_{(2,\, r-1)}+\Lambda _{r}^{4}
\Omega_{(1,\, r-2)}-M_{0}\Omega_{(1,\, r-1)}- \Lambda _{(1,\, r)}^{2}=0.\label{eq:moncona}
\end{equation}
The second set of equations at the second singularity give  $E_{m}=0$ and  (\ref{eq:moncona}) with
 $ \Lambda _{(1,\, r)}^{2}\rightarrow  -\Lambda _{(1,\, r)}^{2} $ and $m\rightarrow d$.

Let us explicitly solve (\ref{eq:moncona}) for $r=2$ and $r=3$. When $r=2$, (\ref{eq:moncona}) with the $\Omega $
functions given in Appendix \ref{sec:ap-om} give
\[
-M_{1}\tilde{E}_{\mathrm{m} }E_{\mathrm{m} }+m_{0}=0,\]
\[
-M_{0}\tilde{E}_{\mathrm{m} }E_{\mathrm{m} }+m_{1}=0,\]
\begin{equation}
\Lambda _{1}^{4}+\Lambda _{2}^{4}-M_{0}M_{1}- 2\Lambda _{1}^{2}\Lambda _{2}^{2}=0.\label{eq:moncon-2a}\end{equation}
The solutions are
\begin{equation}
M_{0}=\epsilon\, (\frac{m_{1}}{m_{0}})^{1/2}(\Lambda _{1}^{2}- \Lambda _{2}^{2}),\quad
M_{1}=\epsilon\, (\frac{m_{0}}{m_{1}})^{1/2}(\Lambda _{1}^{2}- \Lambda _{2}^{2})\label{eq:monconvev-2}\end{equation}
and the expectation values of the monopole condensate \begin{equation}
\tilde{E}_{\mathrm{m} }E_{\mathrm{m} }=\epsilon \, \frac{(m_{0}m_{1})^{\frac{1}{2}}}{\Lambda _{1}^{2}- \Lambda _{2}^{2}},
\label{eq:moncon-2}\end{equation}
where $\epsilon=\pm 1$.
Therefore, the monopole gets confined and a singular submanifold corresponds to
the confining branch of the moduli space.
Note that the above are two solutions at the first singularity. The second set of equations give two more solutions
at the second singularity
with $ \Lambda _{2}^{2}\rightarrow  -\Lambda _{2}^{2} $ in (\ref{eq:moncon-2}).
Thus there are a total of four vacuum states which match the four
phases from gaugino condensation in the low energy pure gauge theory
of two disconnected nodes.

Next consider $r=3$. The equations of motion at the first singularity in this case are
 \[(-M_{1}M_{2}+\Lambda _{2}^{4})\tilde{E}_{\mathrm{m} }E_{\mathrm{m} }+m_{0}=0,\]
\[(-M_{0}M_{2}+\Lambda _{3}^{4})\tilde{E}_{\mathrm{m} }E_{\mathrm{m} }+m_{1}=0,\]
\[(-M_{0}M_{1}+\Lambda _{1}^{4})\tilde{E}_{\mathrm{m} }E_{\mathrm{m} }+m_{2}=0,\]
\begin{equation}
-M_{0}M_{1}M_{2}+\Lambda _{1}^{4}M_{2}+\Lambda _{2}^{4}M_{0}+\Lambda _{3}^{4}M_{1}
-2\Lambda _{1}^{2}\Lambda _{2}^{2}\Lambda _{3}^{2}=0.\label{eq:moncon-3c}
\end{equation}
The same equations of motion as (\ref{eq:moncon-3c}) were obtained in \cite{CEFS} for $r=3$. 
Denoting $\tilde{E}_{\mathrm{m} }E_{\mathrm{m} }$ by $x$, the expectation values
of the condensate are given by the solutions of \begin{eqnarray}
(2\Lambda _{1}^{8}\Lambda _{2}^{4}\Lambda _{3}^{4}m_{0}m_{1}+2\Lambda _{1}^{4}\Lambda _{2}^{4}
\Lambda _{3}^{8}m_{0}m_{2}+2\Lambda _{1}^{4}\Lambda _{2}^{8}\Lambda _{3}^{4}m_{1}m_{2}
-\Lambda _{1}^{8}\Lambda _{2}^{8}m_{1}^{2}-\Lambda _{1}^{8}\Lambda _{3}^{8}m_{0}^{2}
-\Lambda _{2}^{8}\Lambda _{3}^{8}m_{2}^{2})x^{4} &  & \nonumber \\
+\Lambda _{1}^{4}\Lambda _{2}^{4}\Lambda _{3}^{4}m_{0}m_{1}m_{2}x^{3}
+2(\Lambda _{1}^{4}\Lambda _{2}^{4}m_{0}m_{1}^{2}m_{2}+\Lambda _{1}^{4}
\Lambda _{3}^{4}m_{0}^{2}m_{1}m_{2}+\Lambda _{2}^{4}\Lambda _{3}^{4}m_{0}m_{1}m_{2}^{2})x^{2} & & \nonumber \\
-m_{0}^{2}m_{1}^{2}m_{2}^{2}=0. &  & \label{eq:condsol-1}
\end{eqnarray}
Now (\ref{eq:condsol-1}) is a fourth order polynomial equation with four
solutions.
Exactly the same equation holds at the second singularity, since (\ref{eq:condsol-1})
contains even powers of $\Lambda_{i}^{2}$. 
Thus there are a total of eight vacuum
states in the massive theory, four at each singularity. This exactly
matches the eight phases of the low energy theory with all
the matter fields integrated out, where the $Z_{4}\times Z_{4} \times Z_{4}$ $R$
symmetry is broken down to $Z_{2}\times Z_{2}\times Z_{2}$ by gaugino condensation.
Note that $\tilde{E}_{\mathrm{m} }E_{\mathrm{m} }$ in
(\ref{eq:moncon-1}) is a Lagrange multiplier in the language of
Sections \ref{sec:intout-links} - \ref{sec:ring}. In fact, if we
set $\Lambda_{3}\equiv 0$ in (\ref{eq:condsol-1}), we obtain
\begin{equation}
x=(-(\frac{m_{0}m_{2}}{\Lambda _{1}^{4}\Lambda _{2}^{4}})^{\frac{1}{2}},\, -(\frac{m_{0}m_{2}}{
\Lambda _{1}^{4}\Lambda _{2}^{4}})^{\frac{1}{2}},\, (\frac{m_{0}m_{2}}{\Lambda _{1}^{4}\Lambda _{2}^{4}})^{\frac{1}{2}},\,
(\frac{m_{0}m_{2}}{\Lambda _{1}^{4}\Lambda _{2}^{4}})^{\frac{1}{2}})\label{eq:condsol-ls0}
\end{equation}
which exactly matches the non-zero solutions of the linear moose with two nodes given in (\ref{eq:pert-case3x}).  
The extra solution at $x=0$ in (\ref{eq:pert-case3x})
is far out in moduli space.


\setcounter{equation}{0}
\section{Summary\label{summary}}

We started with
simple pure gauge theories of disconnected nodes and 
produced very nontrivial quantum moduli space constraints
and dynamical superpotentials for $\mathcal{N}=1$  $SU(2)^{r}$
 linear and ring moose theories by 
integrating
in and out matter link chiral superfields. We showed that we could
consistently add and remove link fields by exploiting simple and efficient
integrating in and out procedures.
The linear moose with
both external links present has quantum moduli space of vacua.
We have explicitly computed the constraints on the vacua.
We have also shown that when the moduli space is perturbed by 
a generic tree
level superpotential, the vacuum space becomes discrete. The
linear mooses without one or both external links have
non-vanishing low energy dynamical superpotentials. We have
explicitly computed these superpotentials. 
For the ring moose, we found two singular submanifolds with a nontrivial modulus
that is a function of all the
gauge singlets we needed to parameterize the quantum moduli space. 
The massive theory near the 
singularities led to confinement.
The Seiberg-Witten elliptic
curve that describes the quantum moduli space of the ring moose
followed from our computation naturally.

\section*{Acknowledgements}

I am deeply grateful to Howard Georgi for various useful suggestions, comments and discussions.
I am also very thankful to him for sharing his original quantum moduli space relations of the
linear moose with me which initiated this work.  This research is supported in part by
the National Science Foundation under grant number NSF-PHY/98-02709.


\appendix

\setcounter{equation}{0}
\section{\label{sec:ap-om} The $\Omega $ functions}

Here we list the first few $\Omega $ functions defined in (\ref{eq:om-odd})
and (\ref{eq:om-even}).
\begin{eqnarray}
\Omega_{(i,\, i+1)} & = & M_{i}M_{i+1}-\Lambda _{i+1}^{4}\label{eq:ap-omii1}\\
\Omega_{(i,\, i+2)} & = & M_{i}M_{i+1}M_{i+2}-\Lambda _{i+1}^{4}M_{i+2}-\Lambda _{i+2}^{4}M_{i}
\label{eq:ap-omii2}\\
\Omega_{(i,\, i+3)} & = & M_{i}M_{i+1}M_{i+2}M_{i+3}-\Lambda _{i+1}^{4}M_{i+2}M_{i+3}-
\Lambda _{i+2}^{4}M_{i}M_{i+3}\nonumber \\
 &  & -\Lambda _{i+3}^{4}M_{i}M_{i+1}+\Lambda _{i+1}^{4}\Lambda _{i+3}^{4}\label{eq:ap-omii3}\\
\Omega_{(i,\, i+4)} & = & M_{i}M_{i+1}M_{i+2}M_{i+3}M_{i+4}-\Lambda _{i+1}^{4}M_{i+2}M_{i+3}M_{i+4}-
\Lambda _{i+2}^{4}M_{i}M_{i+3}M_{i+4}\nonumber \\
 &  & -\Lambda _{i+3}^{4}M_{i}M_{i+1}M_{i+4}-\Lambda _{i+4}^{4}M_{i}M_{i+1}M_{i+2}\nonumber \\
 &  & +\Lambda _{i+1}^{4}\Lambda _{i+3}^{4}M_{i+4}+\Lambda _{i+1}^{4}\Lambda _{i+4}^{4}M_{i+2}+
\Lambda _{i+2}^{4}\Lambda _{i+4}^{4}M_{i}\label{eq:ap-omii4}\\
\Omega _{(i,\, i+5)} & = & M_{i}M_{i+1}M_{i+2}M_{i+3}M_{i+4}M_{i+5}-\Lambda _{i+1}^{4}M_{i+2}M_{i+3}M_{i+4}M_{i+5}\nonumber \\
 &  & -\Lambda _{i+2}^{4}M_{i}M_{i+3}M_{i+4}M_{i+5}-\Lambda _{i+3}^{4}M_{i}M_{i+1}M_{i+4}M_{i+5}\nonumber \\
 &  & -\Lambda _{i+4}^{4}M_{i}M_{i+1}M_{i+2}M_{i+5}-\Lambda _{i+5}^{4}M_{i}M_{i+1}M_{i+2}M_{i+3}\nonumber \\
 &  & +\Lambda _{i+1}^{4}\Lambda _{i+3}^{4}M_{i+4}M_{i+5}+\Lambda _{i+1}^{4}\Lambda _{i+4}^{4}M_{i+2}M_{i+5}\nonumber \\
 &  & +\Lambda _{i+1}^{4}\Lambda _{i+5}^{4}M_{i+2}M_{i+3}+\Lambda _{i+2}^{4}\Lambda _{i+4}^{4}M_{i}M_{i+5}\nonumber \\
 &  & +\Lambda _{i+2}^{4}\Lambda _{i+5}^{4}M_{i}M_{i+3}+\Lambda _{i+3}^{4}\Lambda _{i+5}^{4}M_{i}M_{i+1}\nonumber \\
 &  & -\Lambda _{i+1}^{4}\Lambda _{i+3}^{4}\Lambda _{i+5}^{4}\label{eq:ap-omii5}
\end{eqnarray}


\setcounter{equation}{0}
\section{\label{sec:ap-om-prop}Some properties of the $\Omega $ functions}

Here we write important recursion relations  we used in our computations that
involve the $\Omega $ functions. We take $j>i$ in all cases unless
explicitly stated.
\begin{eqnarray}
\Omega_{(i,\, i)} & = & M_{i}\label{eq:ap2-omii}\\
\Omega_{(i,\,j)} & = & \Omega_{(i,\, j-1)}M_{j}-\Lambda _{j}^{4}\Omega_{(i,\, j-2)}\label{eq:ap2-omijmj}\\
\Omega_{(i,\,j)} & = & M_{i}\Omega_{(i+1,\, j)}-\Lambda _{i+1}^{4}\Omega_{(i+2,\, j)}\label{eq:ap2-omijmi}\\
\frac{\partial }{\partial M_{k}}\Omega_{(i,\,j)} & = &
\Omega_{(i,\,k-1)}\Omega_{(k+1,\, j)}\label{eq:ap2-domijc}
\end{eqnarray}
\begin{equation}
\Omega_{(i,\, j-2)}\Omega_{(i+1,\, j-1)}-\Omega_{(i,\, j-1)}\Omega_{(i+1,\, j-2)}=\prod _{k=i+1}^{j-1}
\Lambda _{k}^{4}\label{eq:identity-a}\end{equation}
(\ref{eq:ap2-domijc}) for $k=i$ and $k=j$ gives
\begin{eqnarray}
\frac{\partial }{\partial M_{i}}\Omega_{(i,\,j)} & = & \Omega_{(i+1,\, j)},\label{eq:ap2-domij}\\
\frac{\partial }{\partial M_{j}}\Omega_{(i,\,j)} & = & \Omega_{(i,\,
j-1)}.\label{eq:ap2-domijb}
\end{eqnarray}


\setcounter{equation}{0}
\section{\label{sec:ap-pert}Discrete vacua}

Here we will give the discrete vacuum states obtained by minimizing (\ref{eq:superp-quant-2}) with
$M_{0}$, $M_{1}$, $M_{2}$, $T$ and $A$. The equations of motion are
\[
A(-M_{1}M_{2}+\Lambda_{2}^{4})+m_{0}=0,\]
\[
-A\,M_{0}M_{2}+m_{1}=0,\]
\[
A(-M_{0}M_{1}+\Lambda_{1}^{4})+m_{2}=0\]
\[
\mathrm{det}\,T_{(0,\,2)}-M_{0}M_{1}M_{2}+\Lambda _{1}^{4}M_{2}+\Lambda _{2}^{4}M_{0}=0,\]
\begin{equation}
A\,T_{(0,\,2)}^{-1}\,\mathrm{det}\,T_{(0,\,2)}+c=0.
\label{eq:sintout-3ap}\end{equation}
The solution is
\begin{eqnarray}
M_{0} & = & \frac{1}{m_{0}^{2}\,\mathrm{det}\,(c)[\Lambda _{1}^{4}m_{0}-\Lambda _{2}^{4}m_{2}]}
\Bigl (m_{2}\,\mathrm{det}\,(c)^{2}\Lambda _{2}^{8}+m_{0}^{2}m_{1}m_{2}[
\Lambda _{1}^{2}m_{0}-\Lambda _{2}^{4}m_{2}]\nonumber \\
& & +\Lambda _{1}^{4}\Lambda _{2}^{4}[
\mathrm{det}\,(c)^{2}\Lambda _{2}^{4}-m_{0}^{2}m_{1}m_{2}]x
+\Lambda _{1}^{4}\Lambda _{2}^{4}m_{0}m_{1}[-\Lambda _{1}^{4}m_{0}+2
\Lambda _{2}^{4}m_{2}]x^{2}\nonumber \\
& & +\Lambda _{1}^{8}\Lambda _{2}^{8}m_{0}m_{1}x^{3}-
\Lambda _{1}^{8}\Lambda _{2}^{12}m_{1}x^{4}\Bigr ),\label{eq:vevm0-1}\\
M_{1} & = & \frac{1}{\mathrm{det}\,(c)}\Bigl (m_{0}m_{2}-\Lambda _{1}^{4}
\Lambda _{2}^{4}x^{2}\Bigr ),\label{eq:vevm1-1}\\
M_{2} & = & M_{0}(m_{0}\leftrightarrow m_{2},\, \Lambda _{1}^{4}\leftrightarrow
\Lambda _{2}^{4}),\label{eq:vevm2-1}\\
T & = & -\frac{c^{-1}\,x}{\mathrm{det}\,(c)^{2}m_{0}^{2}m_{2}^{2}}
\Bigl [m_{0}^{4}m_{1}^{2}m_{2}^{4}-2m_{0}^{2}m_{1}m_{2}^{2}\,\mathrm{det}\,(c)^{2}[
\Lambda _{1}^{2}m_{0}+\Lambda _{2}^{4}m_{2}]\nonumber \\
& & +\mathrm{det}\,(c)^{4}[\Lambda _{1}^{8}m_{0}^{2}+
\Lambda _{2}^{8}m_{2}^{2}
+\Lambda _{1}^{4}\Lambda _{2}^{4}m_{0}m_{2}]\nonumber \\
& & +\mathrm{det}\,(c)^{2}\Lambda _{1}^{4}
\Lambda _{2}^{4}\Bigl(\mathrm{det}\,(c)^{2}[\Lambda _{1}^{4}m_{0}+
\Lambda _{2}^{4}m_{2}]-3m_{0}^{2}m_{1}m_{2}^{2}\Bigr)x\nonumber \\
 &  & +2\Lambda _{1}^{4}\Lambda _{2}^{4}m_{0}m_{1}m_{2}\Bigl(\mathrm{det}\,(c)^{2}
[\Lambda _{1}^{4}m_{0}+\Lambda _{2}^{4}m_{2}]-m_{0}^{2}m_{1}m_{2}^{2}\Bigr)x^{2}\nonumber \\
 &  & +\mathrm{det}\,(c)^{2}\Lambda _{1}^{8}\Lambda _{2}^{8}m_{0}m_{1}m_{2}x^{3}\nonumber \\
& & -\Lambda _{1}^{8}
\Lambda _{2}^{8}m_{1}\Bigl(\mathrm{det}\,(c)^{2}[\Lambda _{1}^{4}m_{0}+\Lambda _{2}^{4}m_{2}]
 -m_{0}^{2}m_{1}m_{2}^{2}\Bigr)x^{4}\Bigr ],\label{eq:vevt-1}\\
A &= & x, \label{eq:veva-1}
\end{eqnarray}
where
\begin{eqnarray}
 &  & \Lambda _{1}^{8}\Lambda _{2}^{8}m_{1}x^{5}-2\Lambda _{1}^{4}\Lambda _{2}^{4}m_{0}m_{1}m_{2}x^{3}
-\Lambda _{1}^{4}\Lambda _{2}^{4}\textrm{det}(c)^{2}x^{2}\nonumber \\
 &  & -(\Lambda _{1}^{4}m_{0}\textrm{det}(c)^{2}+\Lambda _{2}^{4}m_{2}\textrm{det}(c)^{2}
-m_{0}^{2}m_{1}m_{2}^{2})x-m_{0}m_{2}\textrm{det}(c)^{2}=0.\label{eq:x-1a}
\end{eqnarray}


\setcounter{equation}{0}
\section{\label{sec:ap-wmt}Derivation of (\ref{eq:woneext})}

We want to minimize (\ref{eq:winternalmin}),
\begin{equation}
A\Bigl (\mathrm{det}\,T_{(0,\, r)}-\Omega_{(0,\,r)}\Bigr )+m_{r}M_{r}\label{eq:ap-min}
\end{equation}
with $M_{r}$, $T_{(0,\,r)}$ and $A.$ This gives, using the properties given in Appendix \ref{sec:ap-om-prop},
\begin{equation}
-A\,\Omega_{(0,\, r-1)}+m_{r}=0,\label{eq:ap-dwdmr}\end{equation}
\begin{equation}
A\,T_{(0,\, r)}^{-1}\mathrm{det}\,T_{(0,\, r)}=0,\label{eq:ap-dwdt}\end{equation}
\begin{equation}
\mathrm{det}\,T_{(0,\, r)}-\Omega_{(0,\,r)}=0.\label{eq:dwda}\end{equation}
(\ref{eq:ap-dwdt}) gives \begin{equation}
\mathrm{det}\,T_{(0,\, r)}=0.\label{eq:ap-dwdt-det}\end{equation}
Using (\ref{eq:dwda}) and (\ref{eq:ap-dwdt-det}) with the identity
given in (\ref{eq:identity-a}) in (\ref{eq:dwda}) and solving for $M_{r}$, we get \begin{equation}
M_{r}=\frac{\Lambda _{r}^{4}\Omega_{(0,\, r-2)}}{\Omega_{(0,\, r-1)}}.\label{eq:ap2-mr}\end{equation}
Finally, (\ref{eq:ap2-mr}) and (\ref{eq:dwda}) in (\ref{eq:ap-min}) gives \begin{equation}
W=\frac{\Lambda _{rd}^{5}\Omega_{(0,\, r-2)}}{\Omega_{(0,\, r-1)}}.\label{eq:ap2-wr}\end{equation}


\setcounter{equation}{0}
\section{\label{sec:ap-wm}Derivation of (\ref{eq:winternal})}

Minimizing \begin{equation} W=\frac{\Lambda _{rd}^{5}\Omega_{(0,\,
r-2)}}{\Omega_{(0,\,
r-1)}}+m_{0}M_{0}\label{eq:ap3-wm0}\end{equation} with $M_{0}$ and
using the identity (\ref{eq:ap2-domij}) gives 
\begin{equation}
\Lambda _{rd}^{5}\Omega_{(1,\, r-2)}\Omega_{(0,\, r-1)}-\Lambda
_{rd}^{5}\Omega_{(0,\, r-2)} \Omega_{(1,\, r-1)}+m_{0}\Omega_{(0,\,
r-1)}^{2}=0.\label{eq:ap3-wm1}\end{equation} 
Using the identity (\ref{eq:ap2-omijmi}) in (\ref{eq:ap3-wm1}) gives a quadratic
equation for $M_{0}$,
\begin{eqnarray} \Lambda _{rd}^{5}\Omega_{(1,\,
r-2)}(\Omega_{(1,\, r-1)}M_{0}-\Lambda _{1}^{4}\Omega_{(2,\, r-1)})-
\Lambda _{rd}^{5}(\Omega_{(1,\, r-2)}M_{0} &  & \nonumber \\
-\Lambda _{1}^{4}\Omega_{(2,\, r-2)})\Omega_{(1,\, r-1)}+m_{0}(\Omega_{(1,\, r-1)}M_{0}-\Lambda _{1}^{4}
\Omega_{(2,\, r-1)})^{2}=0 &  & \label{eq:ap3-wm2}
\end{eqnarray}
with solution \begin{eqnarray}
M_{0} & = & \frac{\Lambda _{1d}^{5}\Omega_{(2,\, r-1)}}{m_{0}\Omega_{(1,\, r-1)}}\nonumber \\
 &  & \pm 2\frac{[\Lambda _{1d}^{5}\Lambda _{rd}^{5}(\Omega_{(1,\, r-2)}\Omega_{(2,\, r-1)}-
\Omega_{(1,\, r-1)}\Omega_{(2,\, r-2)})]^{1/2}}{m_{0}\Omega_{(1,\, r-1)}}.\label{eq:ap3-wm3}
\end{eqnarray}
Using (\ref{eq:ap3-wm3}) and the identity (\ref{eq:identity-a})  in
(\ref{eq:ap3-wm0}) gives
\begin{equation}
W=\frac{\Lambda _{1d}^{5}\Omega_{(2,\, r-1)}}{\Omega_{(1,\, r-1)}}+\frac{\Lambda _{rd}^{5}
\Omega_{(1,\, r-2)}}{\Omega_{(1,\, r-1)}}\pm 2\frac{(\Lambda _{1d}^{5}\Lambda _{rd}^{5}
\prod _{i=2}^{r-1}\Lambda _{i}^{4})^{1/2}}{\Omega_{(1,\, r-1)}}.\label{eq:ap3-wm4}
\end{equation}


\setcounter{equation}{0}
\section{\label{sec:ap-ring}Derivation of (\ref{eq:moduli-ring-b})}
From (\ref{eq:wminring})\begin{eqnarray} W & = &
\frac{m_{0}\Lambda _{1}^{4}\Omega_{(2,\, r-1)}}{\Omega_{(1,\,
r-1)}}+\frac{m_{0}\Lambda _{r}^{4} \Omega_{(1,\, r-2)}}{\Omega_{(1,\,
r-1)}}\pm 2\frac{m_{0}(\prod _{i=1}^{r}\Lambda _{i}^{4})^{1/2}}{
\Omega_{(1,\, r-1)}}\nonumber \\
 &  & -\frac{b^{2}}{4m_{0}}\Omega_{(1,\, r-1)}-m_{0}M_{0}-b\,U_{(0,\,r-1)}.\label{eq:ap4-w0}
\end{eqnarray}
Minimizing (\ref{eq:ap4-w0}) with $m_{0}$ and $b$ gives $W=0$
and \begin{eqnarray}
 &  & \frac{\Lambda _{1}^{4}\Omega_{(2,\, r-1)}}{\Omega_{(1,\, r-1)}}+\frac{\Lambda _{r}^{4}
\Omega_{(1,\, r-2)}}{\Omega_{(1,\, r-1)}}\pm 2\frac{(\prod _{i=1}^{r}\Lambda _{i}^{4})^{1/2}}{
\Omega_{(1,\, r-1)}}\nonumber \\
 &  & +\frac{b^{2}}{4m_{0}^{2}}\Omega_{(1,\, r-1)}-M_{0}=0,\label{eq:ap4-w1}
\end{eqnarray}
\begin{equation}
\frac{b}{2m_{0}}=-\frac{U_{(0,\,r-1)}}{\Omega_{(1,\, r-1)}}.\label{eq:ap4-w2}\end{equation}
Putting (\ref{eq:ap4-w2}) in (\ref{eq:ap4-w1}) gives
\begin{equation}
u_{r}\pm \Lambda_{(1,\,r)}^{2}=0,\label{eq:ap4-w3}
\end{equation}
where
\begin{equation}
u_{r}=U_{(0,\,r-1)}^{2}+\Lambda _{1}^{4}\Omega_{(2,\, r-1)}+\Lambda
_{r}^{4}\Omega_{(1,\, r-2)}-M_{0} \Omega_{(1,\, r-1)}\label{eq:ap4-w4}
\end{equation}
and
\begin{equation}
\Lambda_{(1,\,r)}^{2}
=2(\prod _{i=1}^{r}\Lambda _{i}^{4})^{1/2}.\label{eq:ap4-w5}
\end{equation}


\setcounter{equation}{0}
\section{\label{sec:ap-u}The $u$ functions }

Using the definition (\ref{eq:ufun-lam}) or (\ref{eq:ap4-w4}) for
$u_{r}$ and the $\Omega $ functions given in
Appendix \ref{sec:ap-om}, the first few $u$ functions are
\begin{eqnarray}
u_{2} & = & U_{(0,\,1)}^{2}+\Lambda _{1}^{4}+\Lambda _{2}^{4}-M_{0}M_{1},\label{eq:ap4-u2}\\
u_{3} & = & U_{(0,\,2)}^{2}+\Lambda _{1}^{4}M_{2}+\Lambda _{2}^{4}M_{0}+\Lambda _{3}^{4}M_{1}-M_{0}M_{1}M_{2},
\label{eq:ap4-u3}\\
u_{4} & = & U_{(0,\,3)}^{2}+\Lambda _{1}^{4}M_{2}M_{3}+\Lambda _{2}^{4}M_{0}M_{3}+\Lambda _{3}^{4}M_{0}M_{1}+
\Lambda _{4}^{4}M_{1}M_{2}\nonumber \\
 &  & -\Lambda _{1}^{4}\Lambda _{3}^{4}-\Lambda _{2}^{4}\Lambda _{4}^{4}-M_{0}M_{1}M_{2}M_{3},
\label{eq:ap4-u4}\\
u_{5} & = & U_{(0,\,4)}^{2}+\Lambda _{1}^{4}M_{2}M_{3}M_{4}+\Lambda _{2}^{4}M_{0}M_{3}M_{4}+
\Lambda _{3}^{4}M_{0}M_{1}M_{4}+\Lambda _{4}^{4}M_{0}M_{1}M_{2}\nonumber \\
 &  & +\Lambda _{5}^{4}M_{1}M_{2}M_{3}-\Lambda _{1}^{4}\Lambda _{3}^{4}M_{4}-
\Lambda _{1}^{4}\Lambda _{4}^{4}M_{2}-\Lambda _{2}^{4}\Lambda _{5}^{4}M_{3}-
\Lambda _{3}^{4}\Lambda _{5}^{4}M_{1}\nonumber \\
 &  & -\Lambda _{2}^{4}\Lambda _{4}^{4}M_{0}-M_{0}M_{1}M_{2}M_{3}M_{4},
\label{eq:ap4-u5}\\
u_{6} & = & U_{(0,\,5)}^{2}+\Lambda _{1}^{4}M_{2}M_{3}M_{4}M_{5}+\Lambda _{2}^{4}M_{0}M_{3}M_{4}M_{5}
+\Lambda _{3}^{4}M_{0}M_{1}M_{4}M_{5}\nonumber \\
 &  & +\Lambda _{4}^{4}M_{0}M_{1}M_{2}M_{5}+\Lambda _{5}^{4}M_{0}M_{1}M_{2}M_{3}
+\Lambda _{6}^{4}M_{1}M_{2}M_{3}M_{4}\nonumber \\
 &  & -\Lambda _{1}^{4}\Lambda _{3}^{4}M_{4}M_{5}-\Lambda _{1}^{4}\Lambda _{4}^{4}M_{2}M_{5}
-\Lambda _{1}^{4}\Lambda _{5}^{4}M_{2}M_{3}-\Lambda _{2}^{4}\Lambda _{4}^{4}M_{0}M_{5}\nonumber \\
 &  & -\Lambda _{2}^{4}\Lambda _{5}^{4}M_{0}M_{3}-\Lambda _{2}^{4}\Lambda _{6}^{4}M_{3}M_{4}
-\Lambda _{3}^{4}\Lambda _{5}^{4}M_{0}M_{1}-\Lambda _{3}^{4}\Lambda _{6}^{4}M_{1}M_{4}\nonumber \\
 &  & -\Lambda _{4}^{4}\Lambda _{6}^{4}M_{1}M_{2}
+\Lambda _{1}^{4}\Lambda _{3}^{4}\Lambda _{5}^{4}+\Lambda _{2}^{4}\Lambda _{4}^{4}
\Lambda _{6}^{4}-M_{0}M_{1}M_{2}M_{3}M_{4}M_{5}.\label{eq:ap4-u6}
\end{eqnarray}




\begin{thebibliography}{10}

\bibitem{G-1}H. Georgi, {}``A tool kit for builders of composite models,'' \emph{Nucl.
Phys}. \textbf{B266} (1986) 274.
\bibitem{douglas-moore-1}M. R. Douglas and G. Moore, {}``D-branes, quivers, and ALE instantons,''
hep-th/9603167
\bibitem{ACG-1}N. Arkani-Hamed, A. G. Cohen and H. Georgi, {}`` (De)constructing
dimensions,'' \emph{Phys. Rev. Lett}., \textbf{86}, 4757(2001), hep-th/0104005.
\bibitem{HPW-1}C. T. Hill, S. Pokorski, and J. Wang, {}``Gauge invariant effective
Lagrangians for Kaluza-Klein modes,'' \emph{Phys. Rev.} \textbf{D64} (2001) 105005, hep-th/0104035.
\bibitem{ACG-2}N. Arkani-Hamed, A. G. Cohen and H. Georgi, {}``Electroweak symmetry
breaking from dimensional deconstruction,'' \emph{Phys. Lett}., \textbf{B513},
232(2001), hep-ph/0105239; ``Accelerated unification,''
hep-ph/0108089.
\bibitem{CFIKV}F. Cachazo, B. Fiol, K. Intriligator, S. Katz and C. Vafa, {}``A
geometric unification of dualities,'' \emph{Nucl. Phys}. \textbf{B628} (2002)
493;  hep-th/0110028.
\bibitem{seiberg-1}N. Seiberg, {}``Exact results on the space of vacua of four dimensional
susy gauge theories,'' \emph{Phys. Rev.} \textbf{D49} (1994) 6857,
hep-th/9402004.
\bibitem{ILS-1}K. Intriligator, R. G. Leigh and N. Seiberg, {}``Exact superpotentials
in four dimensions,'' \emph{Phys. Rev.} \textbf{D50} (1994) 1092,
hep-th/9403198.
\bibitem{Intriligator-1}K. Intriligator, {}``''Integrating in'' and exact superpotentials
in 4d,'' \emph{Phys. Lett.} \textbf{B336} (1994) 409, hep-th/9407106.
\bibitem{IS-1}K. Intriligator and N. Seiberg, {}``Phases of $\mathcal{N}=1$ supersymmetric
gauge theories in four dimensions,'' \emph{Nucl. Phys}. \textbf{B431}
(1994) 551, hep-th/9408155.
\bibitem{CEFS}C. Csaki, J. Erlich, D. Freedman and W. Skiba, {}``$\mathcal{N}=1$
supersymmetric product group theories in the coulomb phase,'' \emph{Phys.
Rev.} \textbf{D56} (1997) 5209, hep-th/9704067.
\bibitem{ADS}I. Affleck, M. Dine and N. Seiberg, {}``Dynamical supersymmetry breaking
in supersymmetric QCD,'' \emph{Nucl. Phys}. \textbf{B241} (1984)
493; {}``Dynamical supersymmetry breaking in four-dimensions and
its phenomenological implications,'' \emph{Nucl. Phys}. \textbf{B256}
(1985) 557.
\bibitem{SW-1}N. Seiberg and E. Witten, \emph{{}``}Monopole condensation and confinement
in $\mathcal{N}=2$ supersymmetric Yang-Mills theory,'' \emph{Nucl.
Phys}. \textbf{B426} (1994) 19, hep-th/9407081; \emph{{}``}Monopoles,
duality and chiral symmetry breaking in $\mathcal{N}=2$ supersymmetric
QCD,'' \emph{Nucl. Phys}. \textbf{B431} (1994) 484, hep-th/9408099.
\bibitem{S-H}S. Chang and H. Georgi, ``Quantum modified mooses,'' hep-th/0209038.
\bibitem{CEGK}  C. Csaki, J. Erlich, C. Grojean and G. Kribs, ``4D constructions of supersymmetric extra dimensions 
and gaugino mediation.'' \emph{Phys. Rev.} \textbf{D65} (2002) 015003, hep-ph/0106044.
\bibitem{CKT}  C. Csaki, G. D. Kribs and J. Terning, ``4D models of Scherk-Schwarz GUT breaking via 
deconstruction,'' \emph{Phys. Rev.} \textbf{D65} (2002) 015004, hep-ph/0107266.
\bibitem{RS} I. Rothstein  and W. Skiba, ``Mother moose: generating extra dimensions from simple groups at 
large N,'' \emph{Phys. Rev.} \textbf{D65} (2002) 065002, hep-th/0109175.
\bibitem{CEKPSS} C. Csaki, J. Erlich, V. V. Khoze, E. Poppitz, Y. Shadmi, Y. Shirman, ``Exact Results 
in 5D from Instantons and Deconstruction,'' \emph{Phys. Rev.} \textbf{D65} (2002) 085033, hep-th/0110188.

\end{thebibliography}
\end{document}